# BOUCHAUD'S MODEL EXHIBITS TWO DIFFERENT AGING REGIMES IN DIMENSION ONE

By Gérard Ben Arous and Jiří Černý

*Courant Institute of Mathematical Sciences and Weierstrass Institute for Applied Analysis and Stochastics*

Let $E_i$ be a collection of i.i.d. exponential random variables. Bouchaud's model on $\mathbb{Z}$ is a Markov chain $X(t)$ whose transition rates are given by $w_{ij} = \nu \exp(-\beta((1-a)E_i - aE_j))$ if $i$, $j$ are neighbors in $\mathbb{Z}$. We study the behavior of two correlation functions: $\mathbb{P}[X(t_w + t) = X(t_w)]$ and $\mathbb{P}[X(t') = X(t_w) \ \forall\, t' \in [t_w, t_w + t]]$. We prove the (sub)aging behavior of these functions when $\beta > 1$ and $a \in [0, 1]$.

**1. Introduction.** Aging is an out-of-equilibrium physical phenomenon that is gaining considerable interest in contemporary physics and mathematics. An extensive literature exists in physics (see [6] and therein references). The mathematical literature is substantially smaller, although some progress was achieved in recent years ([2, 3, 5, 7, 9]; see also [1] for a survey).

The following model has been proposed by Bouchaud as a toy model for studying the aging phenomenon. Let $G = (\mathcal{V}, \mathcal{E})$ be a graph, and let $E = \{E_i\}_{i \in \mathcal{V}}$ be the collection of i.i.d. random variables indexed by vertices of this graph with the common exponential distribution with mean 1. We consider the continuous-time Markov chain $X(t)$ with state space $\mathcal{V}$, such that

$$(1) \quad \mathbb{P}(X(t+dt) = j | X(t) = i, E) = \begin{cases} w_{ij}\, dt, & \text{if } i, j \text{ are connected in } G, \\ 0, & \text{otherwise.} \end{cases}$$

The transition rates $w_{ij}$ are defined by

$$(2) \qquad\qquad w_{ij} = \nu \exp\big(-\beta((1-a)E_i - aE_j)\big).$$

The parameter $\beta$ denotes, as usual, the inverse temperature and the parameter $a$, $0 \leq a \leq 1$, drives the "symmetry" of the model. The value of $\nu$ fixes the time scale and is irrelevant for our paper; we thus set $\nu = 1$.

This model has been studied when $G$ is $\mathbb{Z}$ and $a = 0$ by Fotnes, Isopi and Newman [8, 9]. It is an elementary model when $G$ is the complete graph, which is a good ansatz for the dynamics of the REM (see [3]).

The time spent by the system at site $i$ grows with the value of $E_i$. The value of $E_i$ can thus be regarded as the depth of the trap at the site $i$. The model is sometimes referred to as "Bouchaud's trap model." It describes the motion of the physical system between the states with energies $-E_i$. It can be regarded as a useful rough approximation of spin-glass dynamics. The states of Bouchaud's trap model correspond to a subset of all possible states of the spin-glass system with exceptionally low energy. This justifies in a certain sense the exponential distribution of $E_i$ since it is the distribution of extreme values. The idea behind this model is that the spin-glass dynamics spends most of the time in the deepest states and it passes through all others extremely quickly. Thus, only the extremal states are important for the long-time behavior of dynamics, which justifies formally the introduction of Bouchaud's model.

Usually, proving an aging result consists in finding a two-point function $F(t_w, t_w + t)$, a quantity that measures the behavior of the system at time $t + t_w$ after it has aged for the time $t_w$, such that a nontrivial limit

$$(3) \qquad \lim_{\substack{t \to \infty \\ t/t_w = \theta}} F(t_w, t_w + t) = F(\theta)$$

exists. The choice of the two-point function is crucial. For instance, it has been observed by Rinn, Maass and Bouchaud [11] that a good choice is

$$(4) \qquad R(t_w, t_w + t) = \mathbb{E}\mathbb{P}(X(t + t_w) = X(t_w)|E),$$

which is the probability that the system will be in the same state at the end of the observation period (i.e., at time $t + t_w$) as it was in the beginning (i.e., at time $t_w$). Another quantity exhibiting aging behavior, which was studied by Fontes, Isopi and Newman [9], is

$$(5) \qquad R^q(t_w, t_w + t) = \mathbb{E} \sum_{i \in \mathbb{Z}} [\mathbb{P}(X(t + t_w) = i|E, X(t_w))]^2,$$

which is the probability that two independent walkers will be at the same site after time $t + t_w$ if they were at the same site at time $t_w$, averaged over the distribution of the common starting point $X(t_w)$. These authors have proved that, for these two two-point functions, aging occurs when $a = 0$. We extend this result to the case $a > 0$. The limiting object will be independent of $a$. Thus the parameter $a$ could seem to be of no relevance for aging.

However, it is not the case for all two-point functions. For instance, for the function

$$(6) \qquad \Pi(t_w, t_w + t) = \mathbb{E}\mathbb{P}(X(t') = X(t_w) \; \forall t' \in [t_w, t_w + t]|E),$$

that is, for the probability that the system does not change its state between $t_w$ and $t_w + t$, it was predicted by Rinn, Maass and Bouchaud [11] that there exists a constant $\gamma$ such that the limit $\lim_{t_w \to \infty} \Pi(t_w, t_w + \theta t_w^\gamma)$ exists and depends nontrivially on $a$. The name *subaging* was introduced for this type of behavior, that is, for the fact that there exists a constant $0 < \gamma < 1$ such that for some two-point function $F(t_w, t_w + t)$, there is a nontrivial limit

$$(7) \qquad\qquad \lim_{\substack{t \to \infty \\ t/t_w^\gamma = \theta}} F(t_w, t_w + t) = F(\theta).$$

One of the main results of the present paper is the proof of the subaging behavior of the function (6) for an arbitrary $a \in [0, 1]$.

Let us have a closer look at the role of the parameter $a$. If $a = 0$, the dynamics of the model is sometimes referred to as "random hopping time (RHT) dynamics" (cf. [10]). In this case the rates $w_{ij}$ do not depend on the value of $E_j$. Hence, the system jumps to all neighboring sites with the same probability and the process $X(t)$ can be regarded as a time change of the simple random walk.

On the other hand, if $a > 0$, the system is attracted to the deepest traps and the underlying discrete-time Markov chain is some kind of random walk in a random environment (RWRE). There are already some results about aging of RWRE in dimension 1 [7]. In that article Sinai's RWRE is considered. It is proved there that there is aging on the scale $\log t / \log t_w \to \text{const}$.

In our situation the energy landscape, far from being seen as a two-sided Brownian motion as in Sinai's RWRE, should be seen as essentially flat with few very narrow deep holes around the deep traps. The drifts on neighboring sites are dependent and this dependency does not allow the existence of large domains with drift in one direction. This can be easily seen by looking at sites surrounding one particularly deep trap $E_i$. Here, the drift at site $i - 1$ pushes the system very strongly to the right and at site $i + 1$ to the left because the system is attracted to the site $i$. Moreover, these drifts have approximately the same size. A more precise description of this picture will be presented later (Section 5). However, these differences do not change notably the mechanism responsible for aging. Again, during the exploration of the random landscape, the process $X$ finds deeper and deeper traps that slow down its dynamics.

It was observed numerically by Rinn, Maass and Bouchaud [11] that $X(t)$ ages only if the temperature is low enough, $\beta > 1$. (In the sequel we will consider only the low-temperature regime.) This heuristically corresponds to the fact that if $a = 0$ and $\beta > 1$, the mean time $\mathbb{E}(\exp(\beta E_0))$ spent by $X(t)$ at an arbitrary site becomes infinite. This implies that the distribution of the depth at which we find the system at time $t$ does not converge as $t \to \infty$. The process $X(t)$ can find deeper and deeper traps where it stays longer.

If $a > 0$, the previous explanation is not precise. The time before the jump is shortened when $a$ increases. On the other hand, the system is attracted to deep traps. This means that instead of staying in one deep trap, the process prefers to jump out and then to return back very quickly. For the two-point functions (4) and (5) these two effects cancel and the limiting behavior is thus independent of $a$. For the two-point function (6), there cannot be cancellation, because the attraction to deeper traps has no influence on it.

Before stating the known results about the model we generalize it slightly. All statements in this paper do not actually require $E_i$ to be an exponential random variable. The only property of $E_i$ that we will need is that the random variable $\exp(\beta E_i)$ is in the domain of attraction of the totally asymmetric stable law with index $\beta^{-1} \equiv \alpha$. Clearly, the original exponential random variable satisfies this property.

Recently, this model was studied rigorously by Fontes, Isopi and Newman [8, 9] in connection with the random voter model and chaotic time dependence. In that papers only the RHT case, $a = 0$, was considered. If $d = 1$ and $\beta > 1$, they proved that the Markov chain $X(t)$ possesses an interesting property called there localization. Namely, it was shown there that

$$\limsup_{t \to \infty} \sup_{i \in \mathbb{Z}} \mathbb{P}(X(t) = i | E) > 0, \qquad \mathbb{P}\text{-a.s.} \tag{8}$$

Also aging for the two-point functions (4) and (5) was proved there. In dimension $d \geq 2$, results of that papers imply that there is no localization in the sense of (8). However, there is numerical evidence [11] that the system ages. A rigorous proof of this claim will be presented in a forthcoming paper [4].

In this article we generalize the results of Fontes, Isopi and Newman [9] in dimension 1 to the general case, $a \neq 0$. As we have already noted, the main difficulty comes from the fact that the underlying discrete-time Markov chain is not a simple random walk. We will prove aging for the quantities (4) and (5). We will then prove subaging for the two-point function (6). As in [9], we relate the asymptotic behavior of quantities (4), (5) and (6) to the similar quantities computed using a singular diffusion $Z(t)$ in a random environment $\rho$—singular meaning here that the single time distributions of $Z$ are discrete.

DEFINITION 1.1 (*Diffusion with random speed measure*).    The random environment $\rho$ is a random discrete measure, $\sum_i v_i \delta_{x_i}$, where the countable collection of $(x_i, v_i)$'s yields an inhomogeneous Poisson point process on $\mathbb{R} \times (0, \infty)$ with density measure $dx \, \alpha v^{-1-\alpha} \, dv$. Conditional on $\rho$, $Z(s)$ is a diffusion process [with $Z(0) = 0$] that can be expressed as a time change of a standard one-dimensional Brownian motion $W(t)$ with the speed measure

$\rho$. Denoting $\ell(t, y)$ the local time of $W(t)$ at $y$, we define

$$\phi^\rho(t) = \int \ell(t, y)\rho(dy) \tag{9}$$

and the stopping time $\psi^\rho(s)$ as the first time $t$ when $\phi^\rho(t) = s$; then $Z(t) = W(\psi^\rho(t))$.

A more detailed description of time changes of Brownian motion can be found in Section 2.

Our main result about aging is the following.

THEOREM 1.2. *For any $\beta > 1$ and $a \in [0, 1]$ there exist nontrivial functions $R(\theta)$, $R^q(\theta)$ such that*

$$\tag{10}\begin{aligned}\lim_{t\to\infty} R(t, t + \theta t) &= \lim_{t\to\infty} \mathbb{E}\mathbb{P}[X((1 + \theta)t) = X(t)|E] = R(\theta),\\ \lim_{t\to\infty} R^q(t, t + \theta t) &= \lim_{t\to\infty} \mathbb{E}\sum_{i\in\mathbb{Z}}\left[\mathbb{P}(X((1 + \theta)t) = i|E, X(t))\right]^2 = R^q(\theta).\end{aligned}$$

*Moreover, $R(\theta)$ and $R^q(\theta)$ can be expressed using the similar quantities defined using the singular diffusion $Z$:*

$$\tag{11}\begin{aligned}R(\theta) &= \mathbb{E}\mathbb{P}[Z(1 + \theta) = Z(1)|\rho],\\ R^q(\theta) &= \mathbb{E}\sum_{x\in\mathbb{R}}\left[\mathbb{P}(Z(1 + \theta) = x|\rho, Z(1))\right]^2.\end{aligned}$$

For $a = 0$, this result is contained in [9]. Since the diffusion $Z(t)$ does not depend on $a$, the functions $R(\theta)$ and $R^q(\theta)$ do not depend on it either. This is the result of the compensation of shorter visits of deep traps by the attraction to them.

We will also prove subaging for the quantity $\Pi(t_w, t_w + t)$. We use $\gamma$ to denote the *subaging exponent*

$$\gamma = \frac{1}{1 + \alpha} = \frac{\beta}{1 + \beta}. \tag{12}$$

THEOREM 1.3. *For any $\beta > 1$ and $a \in [0, 1]$ there exists a nontrivial function $\Pi(\theta)$ such that*

$$\tag{13}\begin{aligned}\lim_{t\to\infty} &\Pi(t, t + f_a(t, \theta))\\ &= \lim_{t\to\infty} \mathbb{E}\mathbb{P}[X(t') = X(t) \ \forall\, t' \in [t, t + f_a(t, \theta)]|E] = \Pi(\theta),\end{aligned}$$

*where the function $f_a$ is given by*

$$f_a(t, \theta) = \theta t^{\gamma(1-a)}L(t)^{1-a}, \tag{14}$$

*and $L(t)$ is a slowly varying function that is determined only by the distribution of $E_0$. Its precise definition is given in Lemma* 8.1. *The function $\Pi(\theta)$ can be again written using the singular diffusion $Z$,*

$$\Pi(\theta) = \int_0^\infty g_a^2(\theta u^{a-1}) \, dF(u), \tag{15}$$

*where $F(u) = \mathbb{E}\mathbb{P}[\rho(Z(1)) \leq u | \rho]$, and where $g_a(\lambda)$ is the Laplace transform $\mathbb{E}(e^{-\lambda T_a})$ of the random variable*

$$T_a = 2^{a-1} \exp(a\beta E_0)[\mathbb{E}(\exp(-2a\beta E_0))]^{1-a}. \tag{16}$$

*If $a = 0$,* (15) *can be written as*

$$\Pi(\theta) = \int_0^\infty e^{-\theta/u} \, dF(u). \tag{17}$$

REMARK. Note that if $E_i$'s are exponential random variables, the function $L(t)$ satisfies $L(t) \equiv 1$. The same is true if $\exp(\beta E_i)$ has a stable law.

As can be seen, in this case the function $\Pi(\theta)$ depends on $a$. This is not surprising since the compensation by attraction has no influence here and the jumps rates clearly depend on $a$.

This behavior of the two-point functions $\Pi(t_w, t + t_w)$ and $R(t_w, t + t_w)$ is not difficult to understand, at least heuristically. One should first look at the behavior of the distribution of the depth of the location of the process at time $t_w$. It can be proved that this depth grows like $t_w^{1/(1+\alpha)}$ (see Proposition 8.2). From this one can see that the main contribution to quantities (4) and (5) comes from trajectories of $X(t)$ that, between times $t_w$ and $t_w + t$, leave $t_w^{(a+\alpha)/(1+\alpha)}$ times the original site and then return to it. Each visit of the original site lasts an amount of time of order $t_w^{(1-a)/(1+\alpha)}$.

In the case of the two-point function (6), we are interested only in the first visit and thus the time $t$ should scale as $t_w^{(1-a)/(1+\alpha)}$. Proofs can be found in Sections 7, 8 and 9. In Section 2 we summarize some known results about time-scale changes of Brownian motion and about point-process convergence. In Section 3 we express the process $X$ and its scaled versions as a time-scale change and in Section 4 we introduce a coupling between the different scales of $X$. In Section 5 we prove convergence of speed measures which is used for time-scale change and we apply this result to show the convergence of finite time distributions of rescaled versions of $X$ to the finite time distributions of $Z$.

**2. Definitions and known results.** In this section we define some notation that we will use often later, and we summarize some known results.

2.1. *Time-scale change of Brownian motion.* The limiting quantities $R(\theta)$, $R^q(\theta)$ and $\Pi(\theta)$ are expressed using the singular diffusion defined by a time change of Brownian motion. So, it will be convenient to express also the chains with discrete state space as a time-scale change of Brownian motion. The scale change is necessary if $a \neq 0$, because the process $X(t)$ does not jump left or right with equal probability.

Consider a locally finite measure

$$(18) \qquad \mu(dx) = \sum_i w_i \delta_{y_i}(dx)$$

which has atoms with weights $w_i$ at positions $y_i$. The measure $\mu$ will be referred to as the speed measure. Let $S$ be a strictly increasing function defined on the set $\{y_i\}$. We call such $S$ the scaling function. Let us introduce slightly nonstandard notation $S \circ \mu$ for the "scaled measure"

$$(19) \qquad (S \circ \mu)(dx) = \sum_i w_i \delta_{S(y_i)}(dx).$$

We use $W(t)$ to denote the standard Brownian motion starting at 0. Let $\ell(t, y)$ be its local time. We define the function

$$(20) \qquad \phi(\mu, S)(t) = \int_{\mathbb{R}} \ell(t, y)(S \circ \mu)(dy)$$

and the stopping time $\psi(\mu, S)(s)$ as the first time when $\phi(\mu, S)(t) = s$. The function $\phi(\mu, S)(t)$ is a nondecreasing, continuous function, and $\psi(\mu, S)(s)$ is its generalized right-continuous inverse. It is an easy corollary of the results of Stone [12] that the process

$$(21) \qquad X(\mu, S)(t) = S^{-1}(W(\psi(\mu, S)(t)))$$

is a time-changed nearest-neighbor random walk on the set of atoms of $\mu$. Moreover, every nearest-neighbor random walk on a countable, nowhere dense subset of $\mathbb{R}$ satisfying some mild conditions on transition probabilities can be expressed in this way. We call the process $X(\mu, S)$ the time-scale change of Brownian motion. If $S$ is the identity function, we speak only about time change.

The following proposition describes the properties of $X(\mu, S)$ if the set of atoms of $\mu$ has no accumulation point. In this case we can suppose that the locations of atoms $y_i$ satisfy $y_i < y_j$ if $i < j$. The claim is the consequence of [12], Section 3. The extra factor 2 comes from the fact that Stone uses the Brownian motion with generator $-\Delta$.

PROPOSITION 2.1. *The process $X(\mu, S)(t)$ is a nearest-neighbor random walk on the set $\{y_i\}$ of atoms of $\mu$. The waiting time in the state $y_i$ is exponentially distributed with mean*

$$(22) \qquad 2w_i \frac{(S(y_{i+1}) - S(y_i))(S(y_i) - S(y_{i-1}))}{S(y_{i+1}) - S(y_{i-1})}.$$

*After leaving state $y_i$, $X(\mu, S)$ enters states $y_{i-1}$ and $y_{i+1}$ with respective probabilities*

$$(23) \qquad \frac{S(y_{i+1}) - S(y_i)}{S(y_{i+1}) - S(y_{i-1})} \quad and \quad \frac{S(y_i) - S(y_{i-1})}{S(y_{i+1}) - S(y_{i-1})}.$$

It will be useful to introduce another process $Y(\mu, S)$ as

$$(24) \qquad Y(\mu, S)(t) = X(S \circ \mu, \mathrm{Id})(t),$$

where Id is the identity function on $\mathbb{R}$. The process $Y(\mu, S)$ can be regarded as $X(\mu, S)$ before the final change of scale in (21). Actually,

$$(25) \qquad Y(\mu, S)(t) = W(\psi(\mu, S)(t)).$$

We will also need processes that are not started at the origin but at some point $x \in \operatorname{supp} \mu$. They are defined in the obvious way using the Brownian motion started at $S(x)$. We use $X(\mu, S; x)$ and $Y(\mu, S; x)$ to denote them.

2.2. *Point-process convergence.* To be able to work with quantities (4)–(6) that have a discrete nature (in the sense that they depend on the probability being exactly at some place) we recall the definition of the point-process convergence of measures introduced in [9]. Let $\mathcal{M}$ denote the set of locally finite measures on $\mathbb{R}$.

DEFINITION 2.2 ([9]). Given a family $\nu$, $\nu^\varepsilon$, $\varepsilon > 0$, in $\mathcal{M}$, we say that $\nu^\varepsilon$ converges *in the point-process sense* to $\nu$, and write $\nu^\varepsilon \overset{pp}{\to} \nu$, as $\varepsilon \to 0$, provided the following holds: if the atoms of $\nu$, $\nu^\varepsilon$ are, respectively, at the distinct locations $y_i$, $y_{i'}^\varepsilon$ with weights $w_i$, $w_{i'}^\varepsilon$, then the subsets of $V^\varepsilon \equiv \bigcup_{i'}\{(y_{i'}^\varepsilon, w_{i'}^\varepsilon)\}$ of $\mathbb{R} \times (0, \infty)$ converge to $V \equiv \bigcup_i \{(y_i, w_i)\}$ as $\varepsilon \to 0$ in the sense that for any open $U$, whose closure $\bar{U}$ is a compact subset of $\mathbb{R} \times (0, \infty)$ such that its boundary contains no points of $V$, the number of points $|V^\varepsilon \cap U|$ in $V^\varepsilon \cap U$ is finite and equals $|V \cap U|$ for all $\varepsilon$ small enough.

Besides this type of convergence we will use the following two more common types of convergence.

DEFINITION 2.3. For the same family as in the previous definition, we say that $\nu^\varepsilon$ converges *vaguely* to $\nu$, and write $\nu^\varepsilon \overset{v}{\to} \nu$, as $\varepsilon \to 0$, if for all continuous real-valued functions $f$ on $\mathbb{R}$ with bounded support $\int f(y)\nu^\varepsilon(dy) \to \int f(y)\nu(dy)$ as $\varepsilon \to 0$. We say that $\nu^\varepsilon$ converges *weakly*, and we write $\nu^\varepsilon \overset{w}{\to} \nu$, as $\varepsilon \to 0$, if the same is true for all bounded continuous functions on $\mathbb{R}$.

To prove the point-process convergence we will use the next lemma which is the copy of Proposition 2.1 of [9].

Let $\nu$, $\nu^\varepsilon$ be locally finite measures on $\mathbb{R}$ and let $(y_i, w_i)$, $(y_i^\varepsilon, w_i^\varepsilon)$ be the sets of atoms of these measures ($y_i$ is the position and $w_i$ is the weight of the atom).

CONDITION 1. *For each $l$ there exists a sequence $j_l(\varepsilon)$ such that*

$$(26) \qquad (y_{j_l(\varepsilon)}^\varepsilon, w_{j_l(\varepsilon)}^\varepsilon) \to (y_l, w_l) \qquad \text{as } \varepsilon \to 0.$$

LEMMA 2.4. *For any family $\nu$, $\nu^\varepsilon$, $\varepsilon > 0$, in $\mathcal{M}$, the following two assertions hold. If $\nu^\varepsilon \xrightarrow{pp} \nu$ as $\varepsilon \to 0$, then Condition 1 holds. If Condition 1 holds and $\nu^\varepsilon \xrightarrow{v} \nu$ as $\varepsilon \to 0$, then also $\nu^\varepsilon \xrightarrow{pp} \nu$ as $\varepsilon \to 0$.*

2.3. *Convergence of the fixed time distributions.* We want to formulate, for future use, a series of results of Fontes, Isopi and Newman [9]. They will allow us to deduce the convergence of fixed time distributions from the convergence of speed measures.

PROPOSITION 2.5. *Let $\mu^\varepsilon$, $\mu$ be the collection of deterministic locally finite measures, and let $Y^\varepsilon$, $Y$ be defined by*

$$(27) \qquad Y^\varepsilon(t) = Y(\mu^\varepsilon, \mathrm{Id})(t) \quad \text{and} \quad Y(t) = Y(\mu, \mathrm{Id})(t).$$

*For any deterministic $t_0 > 0$, let $\nu^\varepsilon$ denote the distribution of $Y^\varepsilon(t_0)$ and let $\nu$ denote the distribution of $Y(t_0)$. Suppose*

$$(28) \qquad \mu^\varepsilon \xrightarrow{v} \mu \quad \text{and} \quad \mu^\varepsilon \xrightarrow{pp} \mu \quad \text{as } \varepsilon \to 0.$$

(i) *Then, as $\varepsilon \to 0$,*

$$(29) \qquad \nu^\varepsilon \xrightarrow{v} \nu \quad \text{and} \quad \nu^\varepsilon \xrightarrow{pp} \nu.$$

(ii) *Let $(x_i^\varepsilon, v_i^\varepsilon)$ and $(x_i, v_i)$ be the collections of atoms of $\mu^\varepsilon$ and $\mu$. Similarly, let $(y_i^\varepsilon, w_i^\varepsilon)$ and $(y_i, w_i)$ be the collections of atoms of $\nu^\varepsilon$ and $\nu$. Then the sets of locations of the atoms are equal,*

$$(30) \qquad \{y_i^\varepsilon\} = \{x_i^\varepsilon\} \quad \text{and} \quad \{y_i\} = \{x_i\}.$$

(iii) *Suppose that we have denoted $x_i$'s and $y_i$'s in such a way that $x_i = y_i$, $x_i^\varepsilon = y_i^\varepsilon$ [which is possible by (ii)]. Let the sequence $j_l(\varepsilon)$ satisfy*

$$(31) \qquad (x_{j_l(\varepsilon)}^\varepsilon, v_{j_l(\varepsilon)}^\varepsilon) \to (x_l, v_l) \qquad \text{as } \varepsilon \to 0.$$

*Then the sequence of corresponding atoms of $\nu^\varepsilon$ satisfies*

$$(32) \qquad (y_{j_l(\varepsilon)}^\varepsilon, w_{j_l(\varepsilon)}^\varepsilon) = (x_{j_l(\varepsilon)}^\varepsilon, w_{j_l(\varepsilon)}^\varepsilon) \to (y_l, w_l) \qquad \text{as } \varepsilon \to 0.$$

(iv) *Let $z^\varepsilon \to z$ and $t^\varepsilon \to t_0$ as $\varepsilon \to 0$. Then parts* (i)–(iii) *stay valid if we replace the process $Y^\varepsilon(t)$ by the process started outside the origin $Y(\mu^\varepsilon, \mathrm{Id}; z^\varepsilon)$, the process $Y(t)$ by $Y(\mu, \mathrm{Id}; z)$, and we define $\nu^\varepsilon$ as the distribution of $Y^\varepsilon(t^\varepsilon)$.*

Part (i) of this proposition is stated as Theorem 2.1 in [9]. Part (ii) is a consequence of Lemmas 2.1 and 2.3 of the same paper. Part (iii) follows from the proof of that theorem, but it is not stated there explicitly. Its proof is, however, the central part of the proof of (i). The remaining part is an easy consequence of (i)–(iii) and of the joint continuity of the local time $\ell(t, y)$.

**3. Expression of $X(t)$ in terms of Brownian motion.** To explore the asymptotic behavior of the chain $X(t)$, we consider its scaling limit

$$X^\varepsilon(t) = \varepsilon X(t / \varepsilon c_\varepsilon). \tag{33}$$

The constant $c_\varepsilon$ will be determined later. For the time being the reader can consider $c_\varepsilon \sim \varepsilon^{1/\alpha}$.

As we already noted in the previous section, it is convenient to express the walks $X(t)$ and $X^\varepsilon(t)$ as a time-scale change of the standard Brownian motion $W(t)$ started at 0. To achieve it we use Proposition 2.1. We define measures

$$\mu(dx) = \mu^1(dx) = \sum_{i \in \mathbb{Z}} \tau_i \delta_i(dx) \quad \text{and} \quad \mu^\varepsilon(dx) = c_\varepsilon \sum_{i \in \mathbb{Z}} \tau_i \delta_{\varepsilon i}(dx), \tag{34}$$

where

$$\tau_i = \tfrac{1}{2} \exp(\beta E_i) \mathbb{E}(\exp(-2a\beta E_0)). \tag{35}$$

We will consider the following scaling function. Let

$$r_i = \frac{\exp(-\beta a(E_i + E_{i+i}))}{\mathbb{E}(\exp(-2\beta a E_0))}, \tag{36}$$

and let

$$S(i) = \begin{cases} \displaystyle\sum_{j=0}^{i-1} r_j, & \text{if } i \geq 0, \\[2ex] \displaystyle -\sum_{j=i}^{-1} r_j, & \text{otherwise.} \end{cases} \tag{37}$$

The constant factor $\mathbb{E}(\exp(-2\beta a E_0))$ that appears in (35) and (36) is not substantial, but it is convenient and it will simplify some expressions later.

We use $\tilde{X}^\varepsilon(t)$, $0 < \varepsilon \leq 1$, to denote the process

$$\tilde{X}^\varepsilon(t) = X(\mu^\varepsilon, \varepsilon S(\varepsilon^{-1} \cdot))(t), \tag{38}$$

which means that $\tilde{X}(t)$ is the time-scale change of Brownian motion with speed measure $\mu^\varepsilon$ and scale function $\varepsilon S(\varepsilon^{-1}\cdot)$. If we write $\psi^\varepsilon(t)$ for $\psi(\mu^\varepsilon, \varepsilon S(\varepsilon^{-1}\cdot))(t)$, we have

$$(39) \qquad \tilde{X}^\varepsilon(t) = \varepsilon S^{-1}(\varepsilon^{-1} W^\varepsilon(\psi^\varepsilon(t))).$$

The process $W^\varepsilon$ is the rescaled Brownian motion, $W^\varepsilon(t) = \varepsilon W(\varepsilon^{-2}t)$, which has the same distribution as $W(t)$. It is introduced only to simplify the proof of the next lemma. In the sequel we will omit the superscript if $\varepsilon = 1$, that is, we will write $\tilde{X}(t)$ for $\tilde{X}^1(t)$, and so on. Note that the function $S^{-1}(\cdot)$ is well defined for all values of its argument. Indeed, the set of atoms of $\varepsilon S(\varepsilon^{-1}\cdot) \circ \mu^\varepsilon$ is the set $\{\varepsilon S(i) : i \in \mathbb{Z}\}$, and thus $\varepsilon^{-1} W^\varepsilon(\psi^\varepsilon(t))$ takes values only in $\{S(i) : i \in \mathbb{Z}\}$.

PROPOSITION 3.1. *The processes $\tilde{X}(t)$ and $\tilde{X}^\varepsilon(t)$ have the same distribution as $X(t)$ and $X^\varepsilon(t) = \varepsilon X(t/c_\varepsilon \varepsilon)$.*

PROOF. We use the symbol "$\sim$" to denote the equality in distribution. The time that $X(t)$ stays at site $i$ is exponentially distributed with mean $(w_{i,i+1} + w_{i,i-1})^{-1}$. The probability that it jumps right or left is

$$(40) \qquad \frac{w_{i,i+1}}{w_{i,i+1} + w_{i,i-1}} \quad \text{and} \quad \frac{w_{i,i-1}}{w_{i,i+1} + w_{i,i-1}}.$$

Plugging the definition (2) of $w_{ij}$ into these expressions, it is easy to see that these values coincide with the same quantities for $\tilde{X}(t)$ which can be computed using Proposition 2.1. This implies that $X(t) \sim \tilde{X}(t)$.

To compare the distributions of $X^\varepsilon(t)$ and $\tilde{X}^\varepsilon(t)$, let us first look at the scaling of $\psi^\varepsilon(t)$. After an easy calculation, using the fact that the local time $\ell^\varepsilon(t, y)$ of $W^\varepsilon$ satisfies $\ell^\varepsilon(t, y) = \varepsilon \ell(\varepsilon^{-2}t, \varepsilon^{-1}y)$, we obtain

$$(41) \qquad \phi^\varepsilon(t) = \int \ell^\varepsilon(t, y)(\varepsilon S(\varepsilon^{-1}\cdot) \circ \mu^\varepsilon)(dy) = \varepsilon c_\varepsilon \phi(\varepsilon^{-2}t).$$

From it we get $\psi^\varepsilon(t) = \varepsilon^2 \psi(t/\varepsilon c_\varepsilon)$. Hence,

$$(42) \qquad \begin{aligned} \varepsilon \tilde{X}(t/\varepsilon c_\varepsilon) &= \varepsilon S^{-1}(W(\psi(t/\varepsilon c_\varepsilon))) = \varepsilon S^{-1}(W(\varepsilon^{-2}\psi^\varepsilon(t))) \\ &= \varepsilon S^{-1}(\varepsilon^{-1} W^\varepsilon(\psi^\varepsilon(t))) = \tilde{X}^\varepsilon(t), \end{aligned}$$

where we used the scaling of $W(t)$ and (39). Since $\tilde{X}(t)$ has the same distribution as $X(t)$, the same is valid for $\tilde{X}^\varepsilon(t)$ and $X^\varepsilon(t)$. $\quad\square$

**4. A coupling for walks on different scales.** It is convenient to introduce the processes $Y(t)$ and $Y^\varepsilon(t)$ that are only a time change of Brownian motion with speed measures $S \circ \mu$ and $\varepsilon S(\varepsilon^{-1} \cdot) \circ \mu^\varepsilon$. Namely,

$$(43) \qquad Y^\varepsilon(t) = Y(\mu^\varepsilon, \varepsilon S(\varepsilon^{-1} \cdot))(t) \quad \text{and} \quad Y(t) = Y(\mu, S)(t).$$

Using (25) we have

$$(44) \qquad Y(t) = W(\psi(t)) \quad \text{and} \quad Y^\varepsilon(t) = W(\psi^\varepsilon(t)).$$

The original processes $X$ and $X^\varepsilon$ are related to them by

$$(45) \qquad X(t) = S^{-1}(Y(t)) \quad \text{and} \quad X^\varepsilon(t) = \varepsilon S^{-1}(\varepsilon^{-1} Y^\varepsilon(t)).$$

In the sequel we want to use Proposition 2.5 to prove the convergence of the finite time distributions of $Y^\varepsilon$. Thus, we want to apply this proposition to the sequence of random speed measures $\mu^\varepsilon$. It is easy to see that convergence in distribution of this sequence is not sufficient for its application. That is why we will construct a coupling between measures $\mu^\varepsilon$ on different scales $\varepsilon$ on a larger probability space. Using this coupling we obtain the a.s. convergence on this space. It is not surprising that the same coupling as in [9] does the job.

Consider the Lévy process $V(x)$, $x \in \mathbb{R}$, $V(0) = 0$, with stationary and independent increments and cadlag paths defined on $(\bar{\Omega}, \bar{\mathcal{F}}, \bar{\mathbb{P}})$ given by

$$(46) \qquad \bar{\mathbb{E}}[e^{ir(V(x+x_0) - V(x_0))}] = \exp\left[x\alpha \int_0^\infty (e^{irw} - 1) w^{-1-\alpha} \, dw\right].$$

Let $\bar{\rho}$ be the random Lebesgue–Stieltjes measure on $\mathbb{R}$ associated to $V$, that is, $\bar{\rho}(a, b] = V(b) - V(a)$. It is a known fact that $\bar{\rho}(dx) = \sum_j v_j \delta_{x_j}(dx)$, where $(x_j, v_j)$ is an inhomogeneous Poisson point process with density $dx \, \alpha v^{-1-\alpha} \, dv$. Note that $\bar{\rho}$ has the same distribution as $\rho$ which we used as speed measure in the definition of the singular diffusion $Z$.

For each fixed $\varepsilon > 0$, we will now define the sequence of i.i.d. random variables $E_i^\varepsilon$ such that $E_i^\varepsilon$'s are defined on the same space as $V$ and $\bar{\rho}$ and they have the same distribution as $E_0$.

Define a function $G : [0, \infty) \mapsto [0, \infty)$ such that

$$(47) \qquad \bar{\mathbb{P}}(V(1) > G(x)) = \mathbb{P}(\tau_0 > x).$$

The function $G$ is well defined since $V(1)$ has continuous distribution, it is nondecreasing and right continuous, and hence has nondecreasing right-continuous generalized inverse $G^{-1}$. Let $g_\varepsilon : [0, \infty) \mapsto [0, \infty)$ be defined as

$$(48) \qquad g_\varepsilon(x) = c_\varepsilon G^{-1}(\varepsilon^{-1/\alpha} x) \qquad \text{for all } x \geq 0,$$

where

$$(49) \qquad c_\varepsilon = (\inf[t \geq 0 : \mathbb{P}(\tau_0 > t) \leq \varepsilon])^{-1}.$$

Note that if $\tau_0$ is the $\alpha$ stable random variable with characteristic function

$$(50) \qquad \mathbb{E}(e^{ir\tau_0}) = \exp\left[\alpha \int_0^\infty (e^{irw} - 1)w^{-1-\alpha}\,dw\right],$$

the choice of $c_\varepsilon$ and $g_\varepsilon$ can be simplified (although it does not correspond to the previous definition):

$$(51) \qquad c_\varepsilon = \varepsilon^{1/\alpha} \quad \text{and} \quad g_\varepsilon(y) \equiv y.$$

The reader who is not interested in the technical details should keep this choice in mind.

LEMMA 4.1. *Let*

$$(52) \qquad \tau_i^\varepsilon = \frac{1}{c_\varepsilon} g_\varepsilon(V(\varepsilon(i+1)) - V(\varepsilon i))$$

*and*

$$(53) \qquad E_i^\varepsilon = \frac{1}{\beta} \log\left(\frac{2\tau_i^\varepsilon}{\mathbb{E}(\exp(-2a\beta E_0))}\right).$$

*Then for any $\varepsilon > 0$, the $\tau_i^\varepsilon$ are i.i.d. with the same law as $\tau_0$, and $\{E_i^\varepsilon\}_{i\in\mathbb{Z}}$ have the same distribution as $\{E_i\}_{i\in\mathbb{Z}}$.*

PROOF. By stationarity and independence of increments of $V$ it is sufficient to show $\bar{\mathbb{P}}(\tau_0^\varepsilon > t) = \mathbb{P}(\tau_0 > t)$. However,

$$(54) \qquad \bar{\mathbb{P}}(\tau_0^\varepsilon > t) = \bar{\mathbb{P}}(V(\varepsilon) > \varepsilon^{1/\alpha} G(t))$$

by the definitions of $\tau_0^\varepsilon$ and $G$. The result then follows from (47) and the scaling invariance of $V$: $V(\varepsilon) \sim \varepsilon^{1/\alpha} V(1)$. The second claim follows easily using (35). □

Let us now define the random speed measures $\bar{\mu}^\varepsilon$ using the collections $\{E_i^\varepsilon\}$ from the previous lemma,

$$(55) \qquad \bar{\mu}^\varepsilon(dx) = \sum_{i\in\mathbb{Z}} c_\varepsilon \tau_i^\varepsilon \delta_{\varepsilon i}(dx).$$

We also define the scaling functions $S_\varepsilon$ similarly as in (37). Let

$$(56) \qquad r_i^\varepsilon = \frac{\exp(-\beta a(E_i^\varepsilon + E_{i+1}^\varepsilon))}{\mathbb{E}(\exp(-2a\beta E_0))}$$

and

$$(57) \qquad S_\varepsilon(i) = \begin{cases} \displaystyle\sum_{j=0}^{i-1} r_j^\varepsilon, & \text{if } i \geq 0, \\[4mm] \displaystyle -\sum_{j=i}^{-1} r_j^\varepsilon, & \text{otherwise.} \end{cases}$$

It is an easy consequence of Lemma 4.1 that $\bar{\mu}^\varepsilon \sim \mu^\varepsilon$ and $S_\varepsilon \sim S$ for any $\varepsilon \in (0,1]$.

**5. Convergence of speed measures.** The following proposition proves the convergence of the scaled speed measures. If $S$ is the identity, that is, $a = 0$, it corresponds to Proposition 3.1 of [9].

PROPOSITION 5.1. *Let $\bar{\mu}^\varepsilon$ and $\bar{\rho}$ be defined as above. Then*

$$\text{(58)} \quad \varepsilon S_\varepsilon(\varepsilon^{-1}\cdot) \circ \bar{\mu}^\varepsilon \xrightarrow{v} \bar{\rho} \quad and \quad \varepsilon S_\varepsilon(\varepsilon^{-1}\cdot) \circ \bar{\mu}^\varepsilon \xrightarrow{pp} \bar{\rho} \qquad as\ \varepsilon \to 0,\ \bar{\mathbb{P}}\text{-}a.s.$$

The proof requires three technical lemmas.

LEMMA 5.2. *As $\varepsilon \to 0$ we have*

$$\text{(59)} \quad \varepsilon S_\varepsilon(\lfloor \varepsilon^{-1} y \rfloor) \to y \qquad as\ \varepsilon \to 0,\ \bar{\mathbb{P}}\text{-}a.s.$$

*uniformly on compact intervals.*

Notice that this lemma sheds more light on the difference between the discrete-time embedded walk of the process $X$ and Sinai's RWRE. In the case of Sinai's RWRE the scale function $S$ corresponds, loosely speaking, to the function

$$\text{(60)} \quad S'(n) = \sum_{i=1}^n \rho_1 \ldots \rho_n,$$

where $\rho_i = (1 - p_i)/p_i$, $p_i$ is the probability going right at $i$, and $p_i$'s are i.i.d. In our case $\rho_i = r_i/r_{i-1}$. An easy computation gives that the product $\rho_1 \ldots \rho_n$ depends only on $E_0$ and $E_{n+1}$. Thus, $S'(n)$ is in our situation essentially a sum of i.i.d. random variables, which is definitively not the case for Sinai's RWRE.

PROOF OF LEMMA 5.2. We consider only $y > 0$. The proof for $y < 0$ is very similar. By definition of $S_\varepsilon$ we have $\varepsilon S_\varepsilon(\lfloor \varepsilon^{-1} y \rfloor) = \varepsilon \sum_{j=0}^{\lfloor \varepsilon^{-1} y \rfloor - 1} r_j^\varepsilon$, where for fixed $\varepsilon$ the sequence $r_j^\varepsilon$ is an ergodic sequence of bounded positive random variables. Moreover, $r_i^\varepsilon$ is independent of all $r_j^\varepsilon$ with $j \notin \{i-1, i, i+1\}$. The $\bar{\mathbb{P}}$-a.s. convergence for fixed $y$ is then a consequence of the strong law of large numbers for triangular arrays. Note that this law of large numbers can be easily proved in our context using the standard methods, because the variables $r_i^\varepsilon$ are bounded and thus their moments of arbitrary large degree are finite. The uniform convergence on compact intervals is easy to prove using the fact that $S_\varepsilon(i)$ is increasing and the identity function is continuous. □

The next two lemmas correspond to Lemmas 3.1 and 3.2 of [9]. We state them without proofs.

LEMMA 5.3.   *For any fixed $y > 0$, $g^\varepsilon(y) \to y$ as $\varepsilon \to 0$.*

LEMMA 5.4.   *For any $\delta' > 0$, there exist constants $C'$ and $C''$ in $(0, \infty)$ such that*

$$(61) \qquad g_\varepsilon(x) \leq C' x^{1-\delta'} \qquad \text{for } \varepsilon^{1/\alpha} \leq x \leq 1 \text{ and } \varepsilon \leq C''.$$

PROOF OF PROPOSITION 5.1.   We first prove the vague convergence. Let $f$ be a bounded continuous function with compact support $I \subset \mathbb{R}$. Then,

$$(62) \quad \int f(x)(\varepsilon S_\varepsilon(\varepsilon^{-1} x) \circ \bar{\mu}^\varepsilon)(dx) = \sum_{i \in J_0^\varepsilon} f(\varepsilon S_\varepsilon(i)) g_\varepsilon(V(\varepsilon(i+1)) - V(\varepsilon i)),$$

where we used the notation

$$(63) \qquad J_y^\varepsilon = \{i \in \mathbb{Z} : \varepsilon S_\varepsilon(i) \in I, V(\varepsilon(i+1)) - V(\varepsilon i) \geq y\}.$$

Choose now $\delta > 0$. To estimate the last sum, we treat separately the sums over $J_\delta^\varepsilon$, $J_{\varepsilon^{1/\alpha}}^\varepsilon \setminus J_\delta^\varepsilon$ and $J_0^\varepsilon \setminus J_{\varepsilon^{1/\alpha}}^\varepsilon$.

Due to the convergence of $\varepsilon S_\varepsilon(\varepsilon^{-1} \cdot)$ to the identity, we know that for $\varepsilon$ small enough there is a small neighborhood $I'$ of $I$ such that $J_0^\varepsilon \subset \varepsilon^{-1} I'$. The process $V$ has $\bar{\mathbb{P}}$-a.s. only finitely many jumps larger than $\delta$ in $I'$, so the first sum has only a finite number of terms. Using the continuity of $f$ and applying Lemmas 5.2 and 5.3 we have

$$(64) \qquad \sum_{i \in J_\delta^\varepsilon} f(\varepsilon S_\varepsilon(i)) g_\varepsilon(V(\varepsilon(i+1)) - V(\varepsilon i)) \to \sum_{j : v_j \geq \delta} f(x_j) v_j,$$

with $(x_i, v_i)$ being the set of atoms of $\bar{\rho}$. In the previous expression we also use the fact that $i\varepsilon \to x_i$ for the corresponding terms in the sums.

By Lemma 5.4 we have for some $\delta'$ such that $\delta' + \alpha \leq 1$

$$
(65) \quad
\begin{aligned}
&\sum_{i \in J_{\varepsilon^{1/\alpha}}^\varepsilon \setminus J_\delta^\varepsilon} f(\varepsilon S_\varepsilon(i)) g_\varepsilon(V(\varepsilon(i+1)) - V(\varepsilon i)) \\
&\leq C \sum_{i \in J_{\varepsilon^{1/\alpha}}^\varepsilon \setminus J_\delta^\varepsilon} (V(\varepsilon(i+1)) - V(\varepsilon i))^{1-\delta'} \leq C \sum_{\substack{j : v_j \leq \delta \\ x_j \in I'}} v_j^{1-\delta'} = H_\delta.
\end{aligned}
$$

From the definition of the point process $(x_i, v_i)$ we have

$$(66) \qquad \bar{\mathbb{E}}(H_\delta) \leq \alpha |I'| \int_0^\delta w^{1-\delta'} w^{-1-\alpha} \, dw \to 0 \qquad \text{as } \delta \to 0.$$

Since $H_\delta$ is decreasing and positive, the limit $\lim_{\delta \to 0} H_\delta$ exists $\bar{\mathbb{P}}$-a.s. The dominated convergence theorem then gives $\bar{\mathbb{E}} \lim_{\delta \to 0} H_\delta = 0$, and thus $\lim_{\delta \to 0} H_\delta = 0$ $\bar{\mathbb{P}}$-a.s.

The third part of the sum is also negligible for $\varepsilon$ small enough. Indeed, by monotonicity of $g_\varepsilon$, we have $g_\varepsilon(x) \leq g_\varepsilon(\varepsilon^{1/\alpha}) \leq C c_\varepsilon$ for all $x \leq \varepsilon^{1/\alpha}$. Hence,

$$
\begin{aligned}
(67) \quad & \sum_{i \in J_0^\varepsilon \setminus J_{\varepsilon^{1/\alpha}}^\varepsilon} f(\varepsilon S_\varepsilon(i)) g_\varepsilon(V(\varepsilon(i+1)) - V(\varepsilon i)) \\
& \leq C' c_\varepsilon \sum_{i \in \varepsilon^{-1} I' \cap \mathbb{Z}} 1 \leq C'' c_\varepsilon \varepsilon^{-1} \to 0 \qquad \text{as } \varepsilon \to 0.
\end{aligned}
$$

In the last equation we use the fact that if $\tau_0$ is in the domain of attraction of the stable law with index $\alpha$, there exists $\kappa > 0$ such that the function $c_\varepsilon$ can be bounded from above by $C \varepsilon^{-\kappa+1/\alpha}$ with $-\kappa + 1/\alpha > 1$.

Putting now all three parts together, we have

$$
\begin{aligned}
(68) \quad & \lim_{\varepsilon \to 0} \sum_{i \in J_0^\varepsilon} f(\varepsilon S_\varepsilon(i)) g_\varepsilon(V(\varepsilon(i+1)) - V(\varepsilon i)) \\
& = \lim_{\delta \to 0} \sum_{j \,:\, v_j \geq \delta} f(x_j) v_j = \int f \, d\bar{\rho}.
\end{aligned}
$$

This proves the vague convergence.

To prove the point-process convergence we use Lemma 2.4. Since we have already proved the vague convergence, we must only verify Condition 1 for the measures $\varepsilon S_\varepsilon(\varepsilon^{-1} \cdot) \circ \bar{\mu}^\varepsilon$ and $\bar{\rho}$. Thus, for any atom $(x_l, v_l)$ of $\bar{\rho}$ we want to find a sequence $j_l(\varepsilon)$ such that

$$
(69) \qquad \varepsilon S_\varepsilon(j_l(\varepsilon)) \to x_l \quad \text{and} \quad g_\varepsilon(V(\varepsilon(j_l(\varepsilon)+1)) - V(\varepsilon j_l(\varepsilon))) \to v_l.
$$

Choose $j_l(\varepsilon)$ such that $x_l \in (\varepsilon j_l(\varepsilon), \varepsilon(j_l(\varepsilon)+1)]$. Then by Lemma 5.2 we have the first statement of (69), and by Lemma 5.3 we have the second. This completes the proof of Proposition 5.1. $\quad \square$

**6. Change of scale for fixed time distributions.** Write $\bar{X}^\varepsilon$ and $\bar{X}$ for the processes defined as in (38), but using the speed measures $\bar{\mu}^\varepsilon$ and the scaling functions $S_\varepsilon$. Since $\bar{\mu}^\varepsilon \sim \mu^\varepsilon$ and $S_\varepsilon \sim S$, we have $\bar{X}^\varepsilon \sim X^\varepsilon$. Similarly, we define the processes $\bar{Y}^\varepsilon$, $\bar{Y}$ as in (44), and $\bar{Z}$ as in Definition 1.1 using the measures with bars. Evidently, $\bar{Y}^\varepsilon \sim Y^\varepsilon$, $\bar{Y} \sim Y$ and $\bar{Z} \sim Z$. The following proposition is a consequence of Propositions 2.5 and 5.1.

PROPOSITION 6.1. *Fix $t_0 > 0$. Write $\bar{\nu}_{Y,V}^\varepsilon$ for the distribution of $\bar{Y}^\varepsilon(t_0)$ and $\bar{\nu}_V$ for the distribution of $\bar{Z}(t_0)$ conditionally on $V$. Then, $\bar{\mathbb{P}}$-a.s. we have*

$$
(70) \qquad \bar{\nu}_{Y,V}^\varepsilon \xrightarrow{v} \bar{\nu}_V \quad \text{and} \quad \bar{\nu}_{Y,V}^\varepsilon \xrightarrow{pp} \bar{\nu}_V \qquad \text{as } \varepsilon \to 0.
$$

The proof of the convergence of the fixed time distribution of $\bar{X}^\varepsilon$ will be finished if we can compare the limits of $\bar{X}^\varepsilon$ and $\bar{Y}^\varepsilon$.

PROPOSITION 6.2. *Fix $t_0$ as in Proposition* 6.1. *Let $\bar{\nu}^\varepsilon_{\bar{X},V}$ denote the distribution of $\bar{X}^\varepsilon(t_0)$ conditionally on $V$. Then, $\bar{\mathbb{P}}$-a.s. we have*

$$\lim_{\varepsilon \to 0} \bar{\nu}^\varepsilon_{\bar{X},V} = \lim_{\varepsilon \to 0} \bar{\nu}^\varepsilon_{\bar{Y},V} = \bar{\nu}_V, \tag{71}$$

*where the limits are taken in both the vague and the point-process sense.*

PROOF. As an easy consequence of Lemma 5.2 we have

$$\varepsilon S_\varepsilon^{-1}(\varepsilon^{-1}y) \to y, \qquad \bar{\mathbb{P}}\text{-a.s.} \tag{72}$$

We will again apply Lemma 2.4 to prove the convergence. Let $f$ be a continuous function with bounded support $I \subset \mathbb{R}$. By continuity of $f$ and (72), choosing the fixed realization of Brownian motion $W$, we have $\bar{\mathbb{P}}$-a.s.

$$\lim_{\varepsilon \to 0} f(\bar{X}^\varepsilon(t_0)) = \lim_{\varepsilon \to 0} f(\bar{Y}^\varepsilon(t_0)). \tag{73}$$

A standard application of the dominated convergence theorem yields

$$\lim_{\varepsilon \to 0} \int f \, d\bar{\nu}^\varepsilon_{\bar{X},V} = \lim_{\varepsilon \to 0} \int f \, d\bar{\nu}^\varepsilon_{\bar{Y},V} = \int f \, d\bar{\nu}_V. \tag{74}$$

We finally verify Condition 1. Write $(x^\varepsilon_i, v^\varepsilon_i)$, $(y^\varepsilon_i, w^\varepsilon_i)$ for the collections of atoms of $\bar{\nu}^\varepsilon_{\bar{X},V}$ and $\bar{\nu}^\varepsilon_{\bar{Y},V}$. By Proposition 2.5(ii) we can choose $x^\varepsilon_i = \varepsilon i$ and $y^\varepsilon_i = \varepsilon S_\varepsilon(i)$, setting eventually $v^\varepsilon_i$, respectively $w^\varepsilon_i$, equal to zero if there is no atom at $x^\varepsilon_i$, respectively $y^\varepsilon_i$. Using this choice of $x^\varepsilon_i$ and $y^\varepsilon_i$ and the relation (45) we have $v^\varepsilon_i = w^\varepsilon_i$. Let $(z_l, u_l)$ be the collection of atoms of $\bar{\nu}_V$ and let $j_l(\varepsilon)$ be the sequence of indexes such that $(y_{j_l(\varepsilon)}, w_{j_l(\varepsilon)}) \to (z_l, u_l)$. Then by (72) we have $(x_{j_l(\varepsilon)}, v_{j_l(\varepsilon)}) \to (z_l, u_l)$ which completes the proof. □

**7. Proof of Theorem 1.2.** We first express the quantities that we are interested in using the processes $\bar{X}^\varepsilon$. From the definition of $\tilde{X}^\varepsilon$, Proposition 3.1 and the fact that $\bar{X}^\varepsilon \sim \tilde{X}^\varepsilon$ we get that if the following limits exist (as we show below), they should satisfy

$$\begin{aligned}
\lim_{t_w \to \infty} \mathbb{E}\mathbb{P}[X((1+\theta)t_w) &= X(t_w)|E] \\
&= \lim_{\varepsilon \to 0} \bar{\mathbb{E}}\bar{\mathbb{P}}[\bar{X}^\varepsilon(1+\theta) = \bar{X}^\varepsilon(1)|V] \\
&\equiv \lim_{\varepsilon \to 0} R_\varepsilon(\theta)
\end{aligned} \tag{75}$$

and similarly

$$\begin{aligned}
\lim_{t_w \to \infty} \mathbb{E} \sum_{i \in \mathbb{Z}} &[\mathbb{P}(X((1+\theta)t_w) = i|E, X(t_w))]^2 \\
&= \lim_{\varepsilon \to 0} \bar{\mathbb{E}} \sum_{i \in \mathbb{Z}} [\bar{\mathbb{P}}(\bar{X}^\varepsilon(1+\theta) = i\varepsilon|V, \bar{X}^\varepsilon(1))]^2 \\
&\equiv \lim_{\varepsilon \to 0} R^q_\varepsilon(\theta).
\end{aligned} \tag{76}$$

We introduce some notation for the sets of atoms of the measures we will consider. In the following everything depends on the realization of the Lévy process $V$ and we will not denote this dependence explicitly. We write

$$(77) \qquad \bar{\mu}^\varepsilon = \sum_i v_i^\varepsilon \delta_{x_i^\varepsilon} \quad \text{and} \quad \bar{\rho} = \sum_i v_i \delta_{x_i}.$$

The atoms of the distribution $\nu_1^\varepsilon$ of $\bar{X}^\varepsilon(1)$ will be denoted by $(x_i^\varepsilon, w_i^\varepsilon)$. Similarly, $(x_i, w_i)$ denotes the atoms of the distribution $\nu_1$ of $\bar{Z}(1)$. The weights of the joint distribution of $\bar{X}^\varepsilon(1)$ and $\bar{X}^\varepsilon(1+\theta)$ will be denoted by $w_{ij}^\varepsilon$,

$$(78) \qquad \begin{aligned} w_{ij}^\varepsilon &= \bar{\mathbb{P}}[(\bar{X}^\varepsilon(1) = x_i^\varepsilon) \cap (\bar{X}^\varepsilon(1+\theta) = x_j^\varepsilon)|V], \\ w_{ij} &= \bar{\mathbb{P}}[(\bar{Z}(1) = x_i) \cap (\bar{Z}(1+\theta) = x_j)|V]. \end{aligned}$$

The last measure we will introduce is the distribution $\nu_{1+\theta}^\varepsilon(\cdot|x_i^\varepsilon)$ of $\bar{X}^\varepsilon(1+\theta)$ conditioned on $\bar{X}^\varepsilon(1) = x_i^\varepsilon$. We denote its atoms by $(x_j^\varepsilon, u_{ij}^\varepsilon)$. Thus,

$$(79) \qquad \begin{aligned} u_{ij}^\varepsilon &= \bar{\mathbb{P}}[\bar{X}^\varepsilon(1+\theta) = x_j^\varepsilon|\bar{X}^\varepsilon(1) = x_i^\varepsilon, V], \\ u_{ij} &= \bar{\mathbb{P}}[\bar{Z}(1+\theta) = x_j|\bar{Z}(1) = x_i, V]. \end{aligned}$$

Observe that $w_{ij}^\varepsilon = w_i^\varepsilon u_{ij}^\varepsilon$ and $w_{ij} = w_i u_{ij}$.

Using this notation we can rewrite (75) and (76),

$$(80) \qquad R_\varepsilon(\theta) = \bar{\mathbb{E}}\left[\sum_i w_i^\varepsilon u_{ii}^\varepsilon\right] \quad \text{and} \quad R_\varepsilon^q(\theta) = \bar{\mathbb{E}}\left[\sum_i w_i^\varepsilon \sum_j (u_{ij}^\varepsilon)^2\right],$$

where the expectations are taken over all realizations of $V$. Obviously we have

$$(81) \qquad R(\theta) = \bar{\mathbb{E}}\left[\sum_i w_i u_{ii}\right] \quad \text{and} \quad R^q(\theta) = \bar{\mathbb{E}}\left[\sum_{i,j} w_i (u_{ij})^2\right].$$

If we prove the $\bar{\mathbb{P}}$-a.s. convergence of the expressions inside the expectations in (80) to the corresponding expressions in (81), the proof will follow easily using the dominated convergence theorem. We want to use the results of Proposition 6.2, namely the point-process convergence of $\nu_1^\varepsilon$ to $\nu_1$ and $\nu_{1+\theta}^\varepsilon(\cdot|x_{j_i(\varepsilon)}^\varepsilon)$ to $\nu_{1+\theta}(\cdot|x_i)$. Here, as usual, $j_i(\varepsilon)$ satisfies $(x_{j_i(\varepsilon)}, v_{j_i(\varepsilon)}) \to (x_i, v_i)$ as $\varepsilon \to 0$. Note that the point-process convergence of $\nu_{1+\theta}^\varepsilon(\cdot|x_{j_i(\varepsilon)}^\varepsilon)$ follows from Propositions 6.2 and 2.5(iv).

In the proof we will need one property of the atoms of different measures that is connected with Condition 1. From the point-process convergence of $\bar{\mu}^\varepsilon$ we know that for every atom $(x_l, v_l)$ of $\bar{\rho}$ there is a function $j_l(\varepsilon)$ such that $(x_{j_l(\varepsilon)}^\varepsilon, v_{j_l(\varepsilon)}^\varepsilon)$ converges to $(x_l, v_l)$. From Proposition 2.5(iii) we can see that for the same function $w_{j_l(\varepsilon)}^\varepsilon \to w_l$, $u_{j_l(\varepsilon), j_k(\varepsilon)}^\varepsilon \to u_{lk}$, and thus $w_{j_l(\varepsilon), j_k(\varepsilon)}^\varepsilon \to$ $w_{lk}$ as $\varepsilon \to 0$. This observation is essential, because only the point-process convergence of all measures is not sufficient to imply our results.

We prove the convergence only for the quantity $R(\theta)$. The proof for $R^q(\theta)$ is entirely similar. Point-process convergence, Condition 1 and the observation of the previous paragraph give

$$(82) \qquad \sum_i w_i u_{ii} = \lim_{\varepsilon \to 0} \sum_i w_{j_i(\varepsilon)}^\varepsilon u_{j_i(\varepsilon), j_i(\varepsilon)}^\varepsilon \leq \liminf_{\varepsilon \to 0} \sum_i w_i^\varepsilon u_{ii}^\varepsilon.$$

To show the opposite bound we choose $\delta > 0$, and divide the sum in (80) into sums over three disjoint sets

$$(83) \qquad \begin{aligned} A_\varepsilon(\delta) &= \{i : w_i^\varepsilon > \delta, u_{ii}^\varepsilon > \delta\}, \\ B_\varepsilon(\delta) &= \{i : u_{ii}^\varepsilon \leq \delta\}, \\ C_\varepsilon(\delta) &= \{i : w_i^\varepsilon \leq \delta, u_{ii}^\varepsilon > \delta\}. \end{aligned}$$

The sum over $A_\varepsilon(\delta)$ has necessarily a finite number of terms. From point-process convergence we have

$$(84) \qquad \limsup_{\varepsilon \to 0} \sum_{i \in A_\varepsilon(\delta)} w_i^\varepsilon u_{ii}^\varepsilon = \sum_{i \in A(\delta)} w_i u_{ii},$$

where $A(\delta)$ has the obvious meaning. For the second part we have

$$(85) \qquad \limsup_{\varepsilon \to 0} \sum_{i \in B_\varepsilon(\delta)} w_i^\varepsilon u_{ii}^\varepsilon \leq \delta \limsup_{\varepsilon \to 0} \sum_{i \in B_\varepsilon(\delta)} w_i^\varepsilon \leq \delta,$$

since $\nu_1^\varepsilon$ is the probability measure. The last part satisfies

$$(86) \qquad \limsup_{\varepsilon \to 0} \sum_{i \in C_\varepsilon(\delta)} w_i^\varepsilon u_{ii}^\varepsilon \leq \limsup_{\varepsilon \to 0} \sum_{i \in C_\varepsilon(\delta)} w_i^\varepsilon \leq 1 - \liminf_{\varepsilon \to 0} \sum_{i \, : \, w_i^\varepsilon > \delta} w_i^\varepsilon.$$

The sum in the last expression has a finite number of terms. Hence

$$(87) \qquad \limsup_{\varepsilon \to 0} \sum_{i \in C_\varepsilon(\delta)} w_i^\varepsilon u_{ii}^\varepsilon \leq 1 - \sum_{i \, : \, w_i > \delta} w_i,$$

and the last sum goes to 1 as $\delta \to 0$, because $\nu_1$ is a purely discrete measure. From (84)–(87) it is easy to see that

$$(88) \qquad \limsup_{\varepsilon \to 0} \sum_i w_i^\varepsilon u_{ii}^\varepsilon \leq \sum_{i \in A(\delta)} w_i u_{ii} + \delta + \left( 1 - \sum_{i \, : \, w_i > \delta} w_i \right)$$

and the proof is completed by taking the limit $\delta \to 0$.

**8. Proof of the sub-aging in symmetric case.** We start the proof by a technical lemma that will provide the connection between the rescaled processes at time $t = 1$ and the process $X$ at some large time $t$. Let $\varepsilon(t)$ be defined by

$$\varepsilon(t) = \sup\{\varepsilon > 0 : \varepsilon c_\varepsilon t \leq 1\}. \tag{89}$$

We write $c_t$ for $c_{\varepsilon(t)}$ and we define $k(t) = \varepsilon(t) c_t t$.

The next lemma defines the slowly varying function $L(t)$ that is used in Theorem 1.3. Note that all slowly varying functions that we use are slowly varying at infinity.

LEMMA 8.1. (i) *There exists a slowly varying function* $L(t)$ *such that*

$$c_t t^\gamma L(t) = 1. \tag{90}$$

(ii) *The function* $k(t)$ *satisfies* $\lim_{t \to \infty} k(t) = 1$.

The proof of this lemma is postponed to the end of the section.

The main step in proving Theorem 1.3 is the following proposition that describes the scaling of the distribution of the depth of the site where $X$ stays at time $t$. We recall that

$$\gamma = \frac{\beta}{1+\beta} = \frac{1}{1+\alpha}. \tag{91}$$

PROPOSITION 8.2. *Let* $F_t(u) = \mathbb{E}\mathbb{P}(\tau(X(t))/t^\gamma L(t) \leq u | E)$. *Then*

$$\lim_{t \to \infty} F_t(u) = \mathbb{E}\mathbb{P}(\rho(Z(1)) \leq u | \rho) \equiv F(u) \tag{92}$$

*for all points of continuity of* $F(u)$.

We use this proposition to prove subaging for $a = 0$.

PROOF OF THEOREM 1.3 IN THE SYMMETRIC CASE. The process $X$ stays at the site $i$ for an exponentially long time with mean $\tau_i$. Using the Markov property we can write

$$\mathbb{P}[X(t') = X(t) \ \forall t' \in [t, t + \theta t^\gamma L(t)]]$$

$$= \int_0^\infty e^{-\theta t^\gamma L(t)/u} \, dF_t(u/(t^\gamma L(t))) = \int_0^\infty e^{-\theta/u} \, dF_t(u). \tag{93}$$

By the weak convergence stated in Proposition 8.2, the last expression converges to $\int e^{-\theta/u} \, dF(u) = \Pi(\theta)$.  □

The proof of Theorem 1.3 for the asymmetric case is postponed to the next section because it is relatively complicated and relies on some notation introduced later in this section.

PROOF OF PROPOSITION 8.2. We follow a strategy similar to that in the proof of aging. Again we start with some notation. Let $h(\varepsilon)$ be such that $\lim_{\varepsilon \to 0} h(\varepsilon) = 1$. We write

$$(94) \qquad \bar{\mu}^{\varepsilon}(dx) = \sum_{i \in \mathbb{Z}} c_{\varepsilon} \tau_i^{\varepsilon} \delta_{i\varepsilon}(dx) \quad \text{and} \quad \bar{\rho}(dx) = \sum_{i \in \mathbb{Z}} v_i \delta_{x_i}(dx).$$

Similarly, the distributions of $\bar{X}^{\varepsilon}(h(\varepsilon))$ and $\bar{Z}(1)$ satisfy

$$(95) \qquad \bar{\nu}_{h(\varepsilon)}^{\varepsilon}(dx) = \sum_{i \in \mathbb{Z}} w_i^{\varepsilon} \delta_{i\varepsilon}(dx) \quad \text{and} \quad \bar{\nu}_1(dx) = \sum_{i \in \mathbb{Z}} w_i \delta_{x_i}(dx).$$

Here again we use the fact that the sets of positions of atoms of $\bar{\rho}$ and $\bar{\nu}_1$ are equal. Note that $w_i^{\varepsilon}$ depends on the function $h$ but we do not denote this dependence explicitly. We also introduce the distributions of the depth at the time $h(\varepsilon)$, respectively, 1:

$$(96) \qquad \pi_{h(\varepsilon)}^{\varepsilon}(dx) = \sum_{i \in \mathbb{Z}} w_i^{\varepsilon} \delta_{c_{\varepsilon} \tau_i^{\varepsilon}}(dx)$$

and

$$(97) \qquad \pi_1(dx) = \sum_{i \in \mathbb{Z}} w_i \delta_{\bar{\rho}(x_i)}(dx) = \sum_{i \in \mathbb{Z}} w_i \delta_{v_i}(dx).$$

We claim that:

LEMMA 8.3.

$$(98) \qquad \pi_{h(\varepsilon)}^{\varepsilon} \overset{v}{\to} \pi_1 \quad \text{and} \quad \pi_{h(\varepsilon)}^{\varepsilon} \overset{pp}{\to} \pi_1 \qquad \text{as } \varepsilon \to 0, \ \bar{\mathbb{P}}\text{-a.s.}$$

PROOF. As usual, we prove the vague convergence and Condition 1. To verify the second property, let us first observe that for any atom $(v_l, w_l)$ of $\pi_1$ there exists $x_l$ such that $(x_l, v_l)$ is an atom of $\bar{\rho}$, and $(x_l, w_l)$ is an atom of $\bar{\nu}_1$. From the point-process convergences $\mu^{\varepsilon} \overset{pp}{\to} \bar{\rho}$, $\bar{\nu}_{h(\varepsilon)}^{\varepsilon} \overset{pp}{\to} \bar{\nu}_1$ and from the direct part of Lemma 2.4 we have that for any $l$ there exist sequences $j_l(\varepsilon)$ and $k_l(\varepsilon)$, such that $(\varepsilon j_l(\varepsilon), c_{\varepsilon} \tau_{j_l(\varepsilon)}^{\varepsilon}) \to (x_l, v_l)$ and $(\varepsilon k_l(\varepsilon), w_{k_l(\varepsilon)}^{\varepsilon}) \to (x_l, w_l)$ as $\varepsilon \to 0$. Moreover, it can be seen from Proposition 2.5(iii) that $j_l(\varepsilon) = k_l(\varepsilon)$. Putting together the last three claims and taking into account that $\bar{\mathbb{P}}$-a.s. $x_l \neq x_m$ implies $v_l \neq v_m$, we easily show that $(c_{\varepsilon} \tau_{j_l(\varepsilon)}^{\varepsilon}, w_{j_l(\varepsilon)}^{\varepsilon}) \to (v_l, w_l)$ as $\varepsilon \to 0$.

We should now verify the vague convergence. Let $f$ be a nonnegative, continuous function with compact support. We use $I_{\delta}$ to denote the open rectangle $(-\delta^{-1}, \delta^{-1}) \times (\delta, 2)$. By (96) we have

$$(99) \qquad \begin{aligned} \int f(x) \pi_{h(\varepsilon)}^{\varepsilon}(dx) &= \sum_{i \in \mathbb{Z}} w_i^{\varepsilon} f(c_{\varepsilon} \tau_{i\varepsilon}^{\varepsilon}) \\ &= \sum_{i:\, (i\varepsilon, w_i^{\varepsilon}) \in I_{\delta}} w_i^{\varepsilon} f(c_{\varepsilon} \tau_{i\varepsilon}^{\varepsilon}) + \sum_{i:\, (i\varepsilon, w_i^{\varepsilon}) \notin I_{\delta}} w_i^{\varepsilon} f(c_{\varepsilon} \tau_{i\varepsilon}^{\varepsilon}). \end{aligned}$$

From the point-process convergence of $\bar{\nu}^\varepsilon_{h(\varepsilon)}$ we know that for all but countably many $\delta > 0$ and for $\varepsilon$ large enough, the number of atoms of $\bar{\nu}^\varepsilon_{h(\varepsilon)}$ in $I_\delta$ is finite and is equal to the number of atoms of $\bar{\nu}_1$ in $I_\delta$. Moreover, by the first part of Lemma 2.4 we have for any such atom $(x_l, w_l)$ the sequence of atoms $(\varepsilon j_l(\varepsilon), w^\varepsilon_{j_l(\varepsilon)})$ converging to $(x_l, w_l)$. By the same reasoning as in the previous paragraph the sequence $c_\varepsilon \tau^\varepsilon_{j_l(\varepsilon)}$ converges as $\varepsilon \to 0$ to $\bar{\rho}(x_l) = v_l$. Thus, by continuity of $f$ we have

$$(100) \qquad \lim_{\varepsilon \to 0} \sum_{i\,:\,(i\varepsilon, w^\varepsilon_i) \in I_\delta} w^\varepsilon_i f(c_\varepsilon \tau^\varepsilon_{i\varepsilon}) = \sum_{i\,:\,(x_i, w_i) \in I_\delta} w_i f(v_i).$$

The right-hand side of the last equation is bounded by $\|f\|_\infty$ and increases as $\delta$ decreases. Thus, its limit as $\delta \to 0$ exists and is equal to $\int f(x)\pi_1(dx)$.

The second sum in (99) is bounded by

$$(101) \qquad C \sum_{i\,:\,(i\varepsilon, w^\varepsilon_i) \notin I_\delta} w^\varepsilon_i = C\left(1 - \sum_{i\,:\,(i\varepsilon, w^\varepsilon_i) \in I_\delta} w^\varepsilon_i\right).$$

Using the same argument as in (87) we have

$$(102) \quad \lim_{\delta \to 0} \limsup_{\varepsilon \to 0}\left(1 - \sum_{i\,:\,(i\varepsilon, w^\varepsilon_i) \in I_\delta} w^\varepsilon_i\right) = \lim_{\delta \to 0}\left(1 - \sum_{i\,:\,(x_i, w_i) \in I_\delta} w_i\right) = 0,$$

since the finite time distribution of $\bar{Z}$ is discrete. $\square$

We can now complete the proof of Proposition 8.2. By definition (33) of $X^\varepsilon(t)$ we have

$$(103) \quad F_t(u) = \mathbb{P}[\tau(X(t))/t^\gamma L(t) \leq u] = \mathbb{P}[\tau(\varepsilon^{-1} X^\varepsilon(t\varepsilon c_\varepsilon))/t^\gamma L(t) \leq u].$$

Inserting the definition (34) of $\mu^\varepsilon$ into the last claim yields

$$(104) \qquad F_t(u) = \mathbb{P}[c_\varepsilon^{-1} \mu^\varepsilon(X^\varepsilon(t\varepsilon c_\varepsilon))/t^\gamma L(t) \leq u].$$

Setting $\varepsilon = \varepsilon(t)$ and using the equality of the distributions $\bar{X}^\varepsilon \sim X^\varepsilon$, $\bar{\mu}^\varepsilon \sim \mu^\varepsilon$ and Lemma 8.1, we get

$$(105) \qquad F_t(u) = \bar{\mathbb{P}}[\bar{\mu}^{\varepsilon(t)}(\bar{X}^{\varepsilon(t)}(k(t))) \leq u].$$

By definition (96) of $\pi^{\varepsilon(t)}_{k(t)}$ we have

$$(106) \quad 1 - F_t(u) = \bar{\mathbb{E}}\bar{\mathbb{P}}[\bar{\mu}^{\varepsilon(t)}(\bar{X}^{\varepsilon(t)}(k(t))) > u | V] = \bar{\mathbb{E}}\left[\sum_{i\,:\,c_t \tau^{\varepsilon(t)}_i > u} w^{\varepsilon(t)}_i\right].$$

The point process convergence proved in Lemma 8.3 implies that the sum in the last expectation converges $\bar{\mathbb{P}}$-a.s. for all $u$ such that $u \neq v_i$ for all $i$:

$$(107) \qquad \lim_{t \to \infty} \sum_{i\,:\,c_t \tau^{\varepsilon(t)}_i > u} w^{\varepsilon(t)}_i = \sum_{i\,:\,v_i > u} w_i = \bar{\mathbb{P}}[\bar{\rho}(\bar{Z}(1)) > u | V].$$

Using the fact that $(\rho, Z)$ has the same distribution as $(\bar{\rho}, \bar{Z})$ and applying the dominated convergence theorem, it is easy to complete the proof. $\square$

PROOF OF LEMMA 8.1. Let $L_1(t)$ be defined by

$$(108) \qquad \mathbb{P}[\tau_0 > t] = t^{-\alpha} L_1(t).$$

Since $\tau_0$ is in the domain of attraction of the stable law with index $\alpha$, the function $L_1$ is slowly varying.

We first show the second claim of the lemma, namely that $k(t) \to 1$ as $t \to \infty$. It is easy to see from (89) that $k(t) \geq 1$. To get an upper bound take $\delta > 0$ and assume that

$$(109) \qquad \limsup_{t \to \infty} k(t) = \limsup_{t \to \infty} \varepsilon(t) c_t t \geq 1 + \delta.$$

If this is true, then there is a sequence $t_n$ such that $t_n \to \infty$ as $n \to \infty$, and $\varepsilon(t_n) c_{t_n} t_n \geq (1 + \delta)$. Using again (89) we get

$$(110) \qquad \varepsilon(t_n) c_{t_n} \geq (1 + \delta) t_n^{-1} \geq (1 + \delta) \lim_{\varepsilon \uparrow \varepsilon(t_n)} \varepsilon c_\varepsilon.$$

This means that $(1 + \delta) c_{\varepsilon(t_n)}^{-1} \leq \lim_{\varepsilon \uparrow \varepsilon(t_n)} c_\varepsilon^{-1}$. Using the definition (49) of $c_\varepsilon$, it is easy to see that this can only happen if there is a sequence $s_n$ such that $s_n \to \infty$ as $n \to \infty$, and $\mathbb{P}[\tau_0 > s_n] = \mathbb{P}[\tau_0 > (1 + \delta) s_n]$. However, then

$$(111) \qquad \frac{L_1((1 + \delta) s_n)}{L_1(s_n)} = \frac{(1 + \delta)^\alpha s_n^\alpha \mathbb{P}[\tau_0 > (1 + \delta) s_n]}{s_n^\alpha \mathbb{P}[\tau_0 > s_n]} = (1 + \delta)^\alpha$$

and this leads to contradiction since $L_1$ is a slowly varying function. Therefore (109) is false and the second part of the lemma is proved.

To verify the first claim of the lemma we should only prove that $L(t)$ is slowly varying. From definition (49) of $c_\varepsilon$ we get

$$(112) \qquad \varepsilon^{-1} \mathbb{P}[\tau_0 > c_\varepsilon^{-1}] \to 1 \qquad \text{as } \varepsilon \to 0.$$

Indeed, it is easy to see that $\varepsilon^{-1} \mathbb{P}[\tau_0 > c_\varepsilon^{-1}] \leq 1$. Take $\eta > 0$; the lower bound follows from

$$(113) \qquad \begin{aligned} (1 + 2\eta)^{-\alpha} &= \lim_{\varepsilon \to 0} \frac{\mathbb{P}[\tau_0 > (1 + 2\eta)/(1 + \eta) c_\varepsilon^{-1}]}{\mathbb{P}[\tau_0 > 1/(1 + \eta) c_\varepsilon^{-1}]} \\ &\leq \liminf_{\varepsilon \to 0} \varepsilon^{-1} \mathbb{P}[\tau_0 > c_\varepsilon^{-1}] \end{aligned}$$

since $\eta$ is arbitrary. From (112) and (108) we get

$$(114) \qquad \varepsilon^{-1} c_\varepsilon^\alpha L_1(c_\varepsilon^{-1}) \to 1 \qquad \text{as } \varepsilon \to 0.$$

Using (114) and $k(t) \to 1$ we get

$$(115) \qquad c_t t^\gamma L_1^\gamma(c_t^{-1}) \to 1 \qquad \text{as } t \to \infty.$$

We want to show that $c_t = t^{-\gamma}L(t)^{-1}$ where $L(t)$ is slowly varying. Choose $k > 0$ and define $d_t = L(t)/L(kt)$. Take $\eta > 0$ small and assume that $\liminf_{t\to\infty} d_t < 1 - 2\eta$. We choose $\delta > 0$ and we consider $t$ large enough such that $c_t t^{\gamma} L_1^{\gamma}(c_t^{-1}) \in (1-\delta, 1+\delta)$. This can be done by (115). We have

$$(116) \quad d_t = \frac{L(t)}{L(kt)} = \frac{c_{kt}}{c_t} k^{\gamma} \geq \frac{1-\delta}{1+\delta} \cdot \frac{L_1^{\gamma}(c_t^{-1})}{L_1^{\gamma}(c_{kt}^{-1})} = \frac{1-\delta}{1+\delta} \cdot \frac{L_1^{\gamma}(c_t^{-1})}{L_1^{\gamma}(d_t^{-1}c_t^{-1}k^{\gamma})}.$$

Our assumption implies that there exists a sequence $t_n$ such that $d_{t_n}^{-1} > 1 + \eta$ for all $n$. Since $L_1$ is slowly varying, we know that for arbitrary $\theta > 0$ there exists $x_0$ such that for all $l > 1 + \eta$ and $x > x_0$ we have $L_1(lx) \leq l^{\theta} L_1(x)$. This implies that for $n$ large enough we have

$$(117) \qquad\qquad d_{t_n} \geq \frac{1-\delta}{1+\delta} \cdot \frac{L_1^{\gamma}(c_{t_n}^{-1})}{d_{t_n}^{-\gamma\theta} L_1^{\gamma}(c_{t_n}^{-1}k^{\gamma})}.$$

Taking the limit $n \to \infty$, using that $c_{t_n} \to \infty$ and that $L_1$ is slowly varying, we get

$$(118) \qquad\qquad \liminf_{n\to\infty} d_{t_n}^{1+\gamma\theta} \geq \frac{1-\delta}{1+\delta}.$$

For every $\eta$ we can take $\delta$ and $\theta$ such that the last equation is in contradiction with $\liminf_{t\to\infty} d_t < 1 - 2\eta$. Thus $\liminf_{t\to\infty} d_t \geq 1$. The proof of the upper bound follows from

$$(119) \qquad\qquad d_{t_n} \leq \frac{1+\delta}{1-\delta} \cdot \frac{d_{t_n}^{\gamma\theta} L_1^{\gamma}(k^{-\gamma}c_{kt_n}^{-1})}{L_1^{\gamma}(c_{kt_n}^{-1})}.$$

This can be proved if one assumes that $\limsup_{t\to\infty} d_t \geq 1 + 2\eta$ and it leads to a contradiction similarly as in (118). □

## 9. Proof of subaging in the nonsymmetric case.

If $a > 0$, the jump rates depend also on the depths of the neighboring sites. As it is easy to see from the definition of $\tau_i^{\varepsilon}$, the depth of the neighboring sites of some very deep trap does not converge $\bar{\mathbb{P}}$-a.s. (By very deep trap we mean here a trap where $X$ has a large chance to stay at time $t$.) On the other hand, we expect (see [11]) that the depth of these sites is, at least if $t_w$ is large, almost independent of the diffusion and has the same distribution as $E_0$. We will show that this expectation is correct.

We consider the function $\Pi(t, t + f_a(t, \theta))$. By its definition we have

$$(120) \begin{aligned} &\Pi(t, t + f_a(t, \theta)) \\ &= \mathbb{E}\left[\sum_{i\in\mathbb{Z}} \mathbb{P}(X(t) = i | E) \exp(-(w_{i,i+1} + w_{i,i-1})f_a(t, \theta))\right]. \end{aligned}$$

The rates $w_{i,i+1}$ and $w_{i,i-1}$ can be expressed using the variables $\tau_i$:

$$(121) \qquad w_{i,i+1} + w_{i,i-1} = \frac{\tau_{i-1}^a + \tau_{i+1}^a}{\tau_i^{1-a}} \left[ \frac{\mathbb{E}(\exp(-2a\beta E_0))}{2} \right]^{1-2a}.$$

We use $K$ to denote the constant in the brackets in the last expression. Then, taking $\varepsilon = \varepsilon(t)$ as in (89),

$$
\begin{aligned}
&\Pi(t, t + f_a(t, \theta)) \\
(122) \qquad &= \mathbb{E}\left[ \sum_{i \in \mathbb{Z}} \mathbb{P}(X^\varepsilon(t \varepsilon c_\varepsilon) = \varepsilon i | E) \exp\left( -K f_a(t, \theta) \frac{\tau_{i+1}^a + \tau_{i-1}^a}{\tau_i^{1-a}} \right) \right] \\
&= \bar{\mathbb{E}}\left[ \sum_{i \in \mathbb{Z}} w_i(t) \exp\left( -K f_a(t, \theta) \frac{(\tau_{i+1}^{\varepsilon(t)})^a + (\tau_{i-1}^{\varepsilon(t)})^a}{(\tau_i^{\varepsilon(t)})^{1-a}} \right) \right],
\end{aligned}
$$

where $w_i(t) = \bar{\mathbb{P}}(\bar{X}^{\varepsilon(t)}(t \varepsilon(t) c_t) = i \varepsilon(t) | V) = \bar{\mathbb{P}}(\bar{X}^{\varepsilon(t)}(k(t)) = i \varepsilon(t) | V)$.

Let $m > 0$ large and $\eta > 0$ small. We use $J_m^\eta = J_m^\eta(V)$ to denote the set of deep traps not far from the origin:

$$(123) \qquad J_m^\eta = \{ x \in [-m, m] : V(x) - V(x-) \geq \eta \}.$$

Let $T_m^\eta(\varepsilon)$ be the set of sites corresponding to $J_m^\eta$ at the scale $\varepsilon$:

$$(124) \qquad T_m^\eta(\varepsilon) = \{ i \in \mathbb{Z} : (i\varepsilon, (i+1)\varepsilon] \cap J_m^\eta \neq \varnothing \}.$$

Note that $J_m^\eta$ and $T_m^\eta(\varepsilon)$ are $\bar{\mathbb{P}}$-a.s. finite sets.

In the following proposition we show that it is possible to choose $m$ and $\eta$ such that $\bar{X}^{\varepsilon(t)}(k(t))$ is with an arbitrarily large probability in $T_m^\eta(\varepsilon(t))$. This can be regarded as a stronger version of the localization effect (8) since the size of the set $T_m^\eta(\varepsilon)$ can be bounded uniformly in $\varepsilon$ by $|J_m^\eta|$.

PROPOSITION 9.1. Let $h(\varepsilon)$ be such that $\lim_{\varepsilon \to 0} h(\varepsilon) = 1$. Then for every $\delta > 0$ there exist $m$, $\eta$ and $\varepsilon_0$ such that for $\varepsilon < \varepsilon_0$

$$(125) \qquad \bar{\mathbb{P}}[\bar{\mathbb{P}}(\varepsilon^{-1} \bar{X}^\varepsilon(h(\varepsilon)) \in T_m^\eta(\varepsilon) | V) > 1 - \delta] > 1 - \delta.$$

We postpone the proof of this proposition to the end of this section and we use it to further simplify (122). Let $\delta > 0$ and let $m$ and $\eta$ be such that (125) holds. We divide the sum in (122) into two parts. The contribution of the sum over $i \notin T_m^\eta(\varepsilon)$ is not important. Indeed, by Proposition 9.1, for all $t$ large enough

$$
\begin{aligned}
(126) \qquad &\bar{\mathbb{E}}\left[ \sum_{i \in \mathbb{Z} \setminus T_m^\eta(\varepsilon(t))} w_i(t) \exp\left( -f_a(t, \theta) K \frac{(\tau_{i+1}^{\varepsilon(t)})^a + (\tau_{i-1}^{\varepsilon(t)})^a}{(\tau_i^{\varepsilon(t)})^{1-a}} \right) \right] \\
&\leq \bar{\mathbb{E}}\left[ \sum_{i \in \mathbb{Z} \setminus T_m^\eta(\varepsilon(t))} w_i(t) \right] \leq 2\delta.
\end{aligned}
$$

To estimate the contribution of the sum over $i \in T_m^\eta(\varepsilon)$ we define the set of neighbors of deep sites

$$(127) \qquad N_m^\eta(\varepsilon) = \{i \in \mathbb{Z} \setminus T_m^\eta(\varepsilon) : \exists j \in T_m^\eta(\varepsilon) \text{ such that } |i - j| = 1\}.$$

Let $\dot\sigma_i^\varepsilon$ be a sequence of i.i.d. random variables defined on $\bar\Omega$ that are independent of $V$ and have the same distribution as $\tau_0^\varepsilon$ conditioned on $J_m^\eta \cap (0, \varepsilon] = \varnothing$. Let $\hat\sigma_i^\varepsilon = \min(\dot\sigma_i^\varepsilon, c_\varepsilon^{-1/2})$. We define

$$(128) \qquad \begin{aligned} \dot\tau_i^\varepsilon &= \begin{cases} \dot\sigma_i^\varepsilon, & \text{for } i \in N_m^\eta(\varepsilon), \\ \tau_i^\varepsilon, & \text{otherwise,} \end{cases} \qquad \text{and} \\[1em] \hat\tau_i^\varepsilon &= \begin{cases} \hat\sigma_i^\varepsilon, & \text{for } i \in N_m^\eta(\varepsilon), \\ \tau_i^\varepsilon, & \text{otherwise.} \end{cases} \end{aligned}$$

We define measures $\dot\mu^\varepsilon$, $\hat\mu^\varepsilon$ and scaling functions $\dot S_\varepsilon$, $\hat S_\varepsilon$ similarly as in (55) and (57) but using $\dot\tau_i^\varepsilon$, $\hat\tau_i^\varepsilon$ instead of $\tau_i^\varepsilon$. Further, let

$$(129) \qquad \dot X^\varepsilon(t) = X(\dot\mu^\varepsilon, \varepsilon\dot S_\varepsilon(\varepsilon^{-1}\cdot))(t) \quad \text{and} \quad \hat X^\varepsilon(t) = X(\hat\mu^\varepsilon, \varepsilon\hat S_\varepsilon(\varepsilon^{-1}\cdot))(t),$$

and let $\dot w_i(t)$, $\hat w_i(t)$ be defined similarly as $w_i(t)$.

To finish the proof of the theorem we will need four technical lemmas.

LEMMA 9.2. *For every fixed realization of $\dot\sigma_i^\varepsilon$, $\bar{\mathbb{P}}$-a.s.*

$$(130) \qquad \varepsilon\hat S_\varepsilon(\varepsilon^{-1}\cdot) \circ \hat\mu^\varepsilon \xrightarrow{v} \bar\rho \quad \text{and} \quad \varepsilon\hat S_\varepsilon(\varepsilon^{-1}\cdot) \circ \hat\mu^\varepsilon \xrightarrow{pp} \bar\rho \qquad \text{as } \varepsilon \to 0.$$

*Therefore, the distribution of $\hat X^{\varepsilon(t)}(k(t))$ converges as $t \to \infty$ weakly and in the point-process sense to the distribution of $\bar Z(1)$.*

*In particular, for all $x \in J_m^\eta$*

$$(131) \qquad \lim_{t \to \infty} \hat w_{j_{\varepsilon(t)}}(t) = \bar{\bar{\mathbb{P}}}(\bar Z(1) = x | V) \equiv w_x \quad \text{and} \quad \lim_{\varepsilon \to 0} c_\varepsilon \hat\tau_{j_\varepsilon}^\varepsilon = \bar\rho(x),$$

*where $j_\varepsilon = j_\varepsilon(x) \in T_m^\eta(\varepsilon)$ satisfies $x \in (\varepsilon j_\varepsilon, \varepsilon(j_\varepsilon + 1)]$.*

PROOF. The proof of the first part of this lemma is very similar to the proofs of Lemma 5.2 and Proposition 5.1; the finite number of changes of neighbors of the deep traps loses its influence as $\varepsilon \to 0$. Therefore, we only describe modifications that must be done in the original proofs.

To get an equivalent of Lemma 5.2 we must show that $\varepsilon\hat S_\varepsilon(\lfloor\varepsilon^{-1}y\rfloor) = \varepsilon\sum_{j=0}^{\lfloor\varepsilon^{-1}y\rfloor}\hat\tau_j^\varepsilon$ converges to $y$. Since $J_m^\eta$ is finite, only a finite number of $\hat\tau_j^\varepsilon$'s is influenced by changing the sequence of $\tau$'s. Since $\hat\tau_j^\varepsilon$ are bounded, the contribution of the changed part of the sum tends to zero as $\varepsilon \to 0$. The rest of the sum can be treated in the same way as in the proof of Lemma 5.2.

Further, we must show the vague convergence and Condition 1 for the measures $\varepsilon\hat S_\varepsilon(\varepsilon^{-1}\cdot) \circ \hat\mu^\varepsilon$. Let $x$ be position of an atom of $\rho$ and let $i(\varepsilon)$ be given by $x \in (\varepsilon i(\varepsilon), \varepsilon(i(\varepsilon) + 1)]$. It is easy to observe that $i(\varepsilon) \notin N_m^\eta(\varepsilon)$ for all $\varepsilon$ small enough. The proof of Condition 1 can be then finished using the same reasoning as before.

To show the vague convergence let $f(x)$ be a bounded continuous function with bounded support. Then

$$
(132) \quad \left| \int f(x)(\varepsilon \hat{S}_\varepsilon(\varepsilon^{-1}\cdot) \circ \hat{\mu}^\varepsilon)(dx) - \int f(x)(\varepsilon S_\varepsilon(\varepsilon^{-1}\cdot) \circ \bar{\mu}^\varepsilon)(dx) \right|
$$

$$
\leq \left| \sum_{i \in N_m^\eta(\varepsilon)} f(\varepsilon S_\varepsilon(i)) g_\varepsilon(\bar{\rho}(i\varepsilon, (i+1)\varepsilon)) \right| + \left| \sum_{i \in N_m^\eta(\varepsilon)} f(\varepsilon \hat{S}_\varepsilon(i)) c_\varepsilon \hat{\sigma}_i^\varepsilon \right|.
$$

The contribution of the first term can be proved to be small observing that $J_\delta^\varepsilon$ [defined in (63)] satisfies $J_\delta^\varepsilon \cap N_m^\eta(\varepsilon) = \varnothing$ for all $\delta > 0$ if $\varepsilon$ is small enough. The second term in (132) is also negligible since $\hat{\sigma}_i^\varepsilon \leq c_\varepsilon^{-1/2}$ and $|N_m^\eta(\varepsilon)| \leq 2|J_m^\eta|$ is a.s. finite.

The convergence of $\hat{X}^{\varepsilon(t)}(k(t))$ and of $w_{j_{\varepsilon(t)}}(t)$ is then a consequence of the first part of the lemma and Proposition 2.5. □

LEMMA 9.3. *The sequence $\dot{\tau}_i^\varepsilon$ has the same distribution as $\tau_i$.*

PROOF. The proof is obvious because the distribution of $\dot{\sigma}_i^\varepsilon$ is chosen to be equal to the distribution of $\tau_i^\varepsilon$ conditioned on $i \notin T_m^\eta(\varepsilon)$. □

LEMMA 9.4. *For $\bar{\mathbb{P}}$-a.e. realization of $V$*

$$
(133) \quad \lim_{\varepsilon \to 0} \bar{\mathbb{P}}(\exists i \in N_m^\eta(\varepsilon) : \dot{\sigma}_i^\varepsilon \neq \hat{\sigma}_i^\varepsilon | V) = 0.
$$

PROOF. The probability that $\dot{\sigma}_i^\varepsilon \geq c_\varepsilon^{-1/2}$ tends to zero. Since $N_m^\eta(\varepsilon)$ is a.s. finite, the proof is complete. □

LEMMA 9.5. *As $\varepsilon \to 0$ the random variables $\hat{\sigma}_0^\varepsilon$ converge weakly to $\tau_0$.*

PROOF. By definition of $\dot{\sigma}_i$, $\bar{\mathbb{P}}(\dot{\sigma}_0^\varepsilon \leq a) = \bar{\mathbb{P}}(\tau_0^\varepsilon \leq a | J_m^\eta \cap (0, \varepsilon] = \varnothing)$. Since the probability of the conditioning event tends to 1 as $\varepsilon \to 0$ and $\tau_0^\varepsilon$ has the same distribution as $\tau_0$, this converges to $\bar{\mathbb{P}}(\tau_0 \leq a)$. Therefore, $\dot{\sigma}_0^\varepsilon$ converges weakly to $\tau_0$. Since $\hat{\sigma}_0^\varepsilon = \min(\dot{\sigma}_0^\varepsilon, c_\varepsilon^{-1/2})$ and $c_\varepsilon^{-1/2} \to \infty$ as $\varepsilon \to 0$, the lemma follows.

□

We can now estimate the contribution of the sum over $i \in T_m^\eta(\varepsilon)$ in (122). Using Lemma 9.3 we get

$$
\bar{\mathbb{E}}\left[ \sum_{i \in T_m^\eta(\varepsilon(t))} w_i(t) \exp\left( -K f_a(t, \theta) \frac{(\tau_{i+1}^{\varepsilon(t)})^a + (\tau_{i-1}^{\varepsilon(t)})^a}{(\tau_i^{\varepsilon(t)})^{1-a}} \right) \right]
$$

$$
\begin{aligned}
(134) \qquad &= \bar{\bar{\mathbb{E}}}\Bigg[\sum_{i \in T^\eta_m(\varepsilon(t))} \dot{w}_i(t) \exp\Bigg(-Kf_a(t,\theta)\frac{(\hat{\tau}^{\varepsilon(t)}_{i+1})^a + (\hat{\tau}^{\varepsilon(t)}_{i-1})^a}{(\hat{\tau}^{\varepsilon(t)}_i)^{1-a}}\Bigg)\Bigg] \\
&= \bar{\bar{\mathbb{E}}}\Bigg[\sum_{i \in T^\eta_m(\varepsilon(t))} \hat{w}_i(t) \exp\Bigg(-Kf_a(t,\theta)\frac{(\hat{\tau}^{\varepsilon(t)}_{i+1})^a + (\hat{\tau}^{\varepsilon(t)}_{i-1})^a}{(\hat{\tau}^{\varepsilon(t)}_i)^{1-a}}\Bigg)\Bigg] \\
&\quad + R_1(\varepsilon(t)).
\end{aligned}
$$

The error term $R_1(\varepsilon)$ can be bounded by

$$
(135) \qquad |R_1(\varepsilon)| \le \bar{\bar{\mathbb{E}}}[\bar{\mathbb{P}}(\exists\, i \in N^\eta_m(\varepsilon) : \dot{\sigma}^\varepsilon_i \ne \hat{\sigma}^\varepsilon_i | V)] \to 0 \qquad \text{as } \varepsilon \to 0
$$

by Lemma 9.4 and the dominated convergence theorem. Recall that $f_a(t,\theta) = \theta t^{\gamma(1-\alpha)} L(t)^{1-\alpha}$. Therefore, using Lemma 8.1, the main term in (134) can be rewritten as

$$
\begin{aligned}
(136) \qquad &\bar{\bar{\mathbb{E}}}\Bigg[\sum_{x \in J^\eta_m} w_x \exp\Bigg(-Kf_a(t,\theta)\frac{(\hat{\tau}^{\varepsilon(t)}_{j_{\varepsilon(t)}(x)+1})^a + (\hat{\tau}^{\varepsilon(t)}_{j_{\varepsilon(t)}(x)-1})^a}{c_t^{a-1}\bar{\rho}(x)^{1-a}}\Bigg)\Bigg] + R_2(\varepsilon(t)) \\
&= \bar{\bar{\mathbb{E}}}\Bigg[\sum_{x \in J^\eta_m} w_x \exp\Bigg(-K\theta\frac{(\hat{\sigma}^{\varepsilon(t)}_{j_{\varepsilon(t)}(x)+1})^a + (\hat{\sigma}^{\varepsilon(t)}_{j_{\varepsilon(t)}(x)-1})^a}{\bar{\rho}(x)^{1-a}}\Bigg)\Bigg] + R_2(\varepsilon(t)),
\end{aligned}
$$

where $j_\varepsilon(x)$ is defined as in Lemma 9.2 and $R_2(\varepsilon)$ is an error that comes from the replacement of $\hat{w}_i(t)$ and $\hat{\tau}^{\varepsilon(t)}_i$ by $w_x$ and $c_t\bar{\rho}(x)$. It follows from Lemma 9.2 that $|R_2(\varepsilon)| \to 0$ as $\varepsilon \to 0$.

We can now easily compute the expectation over $\hat{\sigma}$ in (136). Let $g^\varepsilon_a(\lambda)$ denote the Laplace transform of $K(\hat{\sigma}^\varepsilon_0)^a$, $g^\varepsilon_a(\lambda) = \bar{\bar{\mathbb{E}}}(\exp(-\lambda K(\hat{\sigma}^\varepsilon_0)^a))$, and let $g_a(\lambda) = \mathbb{E}(\exp(-\lambda K\tau^a_0))$. Since $\tau_0$ has the same distribution as $\exp(\beta E_0) \times \mathbb{E}(\exp(-2a\beta E_0))/2$, $K\tau^a_0$ has the same distribution as

$$
(137) \qquad 2^{a-1}\exp(a\beta E_0)(\mathbb{E}(\exp(-2a\beta E_0)))^{1-a} \equiv T_a,
$$

and $g_a(\lambda) = \mathbb{E}(e^{-\lambda T_a})$ as required by Theorem 1.3. From Lemma 9.5 it follows that $\lim_{\varepsilon \to 0} g^\varepsilon_a(\lambda) = g_a(\lambda)$. Using this notation, (126), (134) and (136) we get

$$
\begin{aligned}
&\limsup_{t \to \infty} \Pi(t, f_a(t,\theta)) \\
(138) \qquad &\le \limsup_{\varepsilon \to 0} \bar{\bar{\mathbb{E}}}\Bigg[\sum_{x \in J^\eta_m} w_x \exp\Bigg(-K\theta\frac{(\hat{\sigma}^\varepsilon_{j_\varepsilon(x)+1})^a + (\hat{\sigma}^\varepsilon_{j_\varepsilon(x)-1})^a}{\bar{\rho}(x)^{1-a}}\Bigg)\Bigg] + 2\delta \\
&= \bar{\bar{\mathbb{E}}}\Bigg[\sum_{x \in J^\eta_m} w_x g^2_a(\theta\bar{\rho}(x)^{a-1})\Bigg] + 2\delta.
\end{aligned}
$$

Inserting the remaining atoms of $\bar{\rho}$ inside the sum, making again an error of order at most $2\delta$, we get

$$(139) \qquad \limsup_{t \to \infty} \Pi(t, f_a(t, \theta)) \leq \bar{\mathbb{E}}\left[\sum_x w_x g_a^2(\theta \bar{\rho}(x)^{a-1})\right] + 4\delta.$$

An analogous calculation gives

$$(140) \qquad \liminf_{t \to \infty} \Pi(t, f_a(t, \theta)) \geq \bar{\mathbb{E}}\left[\sum_x w_x g_a^2(\theta \bar{\rho}(x)^{a-1})\right] - 4\delta.$$

Since $\delta$ was arbitrary we have

$$(141) \qquad \Pi(\theta) = \int_0^\infty g_a^2(\theta u^{a-1}) \, dF(u),$$

which finishes the proof of subaging in the asymmetric situation. We still have to show Proposition 9.1.

PROOF OF PROPOSITION 9.1. The claim follows from the existence of $\eta$ and $m$ such that

$$(142) \qquad \bar{\mathbb{P}}[\bar{\mathbb{P}}(\bar{Z}(1) \in J_m^\eta | V) \geq 1 - \delta/2] \geq 1 - \delta/2,$$

and from the $\bar{\mathbb{P}}$-a.s. point-process convergence of the distribution of $\hat{X}^\varepsilon(1)$ to that of $\bar{Z}(1)$. Namely, for $\bar{\mathbb{P}}$-a.e. realization of $V$ it follows from Proposition 2.5(iii), (iv) that there is $\varepsilon(V) > 0$ such that for $\varepsilon < \varepsilon(V)$

$$(143) \qquad |\bar{\mathbb{P}}(\bar{Z}(1) \in J_m^\eta | V) - \bar{\mathbb{P}}(\bar{X}^\varepsilon(h(\varepsilon)) \in T_m^\eta(\varepsilon) | V)| \leq \delta/2.$$

We then take $\varepsilon_0$ such that $\bar{\mathbb{P}}(\varepsilon(V) > \varepsilon_0) > 1 - \delta/2$.

We should still verify (142). It is equivalent to

$$(144) \qquad \bar{\mathbb{P}}[\bar{\mathbb{P}}(\bar{Z}(1) \notin J_m^\eta | V) \leq \delta/2] \geq 1 - \delta/2.$$

The last claim can be easily verified if we show

$$(145) \qquad \bar{\mathbb{P}}[\bar{Z}(1) \notin J_m^\eta] = \bar{\mathbb{E}}[\bar{\mathbb{P}}(\bar{Z}(1) \notin J_m^\eta | V)] \leq \delta^2/4.$$

Indeed, assume that (144) is not true, that is,

$$(146) \qquad \bar{\mathbb{P}}[\bar{\mathbb{P}}(\bar{Z}(1) \notin J_m^\eta | V) > \delta/2] > \delta/2.$$

Then clearly

$$(147) \qquad \bar{\mathbb{E}}[\bar{\mathbb{P}}(\bar{Z}(1) \notin J_m^\eta | V)] > \delta^2/4,$$

in contradiction with (145).

We establish claim (145) using two lemmas.

LEMMA 9.6.    *Let $\eta(t) = t^{1/(1+\alpha)}$ and $m(t) = t^{\alpha/(1+\alpha)}$. Then*

$$\bar{\mathbb{P}}(\bar{Z}(1) \in J_m^\eta) = \bar{\mathbb{P}}(\bar{Z}(t) \in J_{m(t)}^{\eta(t)}). \tag{148}$$

LEMMA 9.7.    *For every $\delta'$ there exist $m'$ and $\eta'$ such that*

$$\int_0^1 \bar{\mathbb{P}}(\bar{Z}(t) \in J_{m'}^{\eta'}) \, dt \geq 1 - \delta'. \tag{149}$$

We first finish the proof of Proposition 9.1. Lemma 9.7 ensures the existence of $t \in (0,1)$ such that $\bar{\mathbb{P}}(Z(t) \in J_{m'}^{\eta'}) \geq 1 - \delta'$. The claim (145) then follows from Lemma 9.6, choosing $\delta' = \delta^2/4$, $m = t^{-\alpha/(1+\alpha)}m'$ and $\eta = t^{-1/(1+\alpha)}\eta'$. $\square$

PROOF OF LEMMA 9.6.    The pair

$$(W_\lambda(t), V_\lambda(x)) \equiv (\lambda W(\lambda^{-2}t), \lambda^{1/\alpha}V(\lambda^{-1}x)) \tag{150}$$

has the same distribution as $(W(t), V(x))$. The measure $\bar{\rho}_\lambda$ associated to $V_\lambda$ can be written as

$$\bar{\rho}_\lambda = \sum_{x_i}(V_\lambda(x_i) - V_\lambda(x_i-))\delta_{x_i} = \lambda^{1/\alpha}\sum_{y_i}(V(y_i) - V(y_i-))\delta_{\lambda y_i}. \tag{151}$$

We thus have

$$\begin{aligned}
\phi_\lambda(t) &\equiv \int \ell_\lambda(t,y)\bar{\rho}_\lambda(dy) = \int \lambda\ell(\lambda^{-2}t, \lambda^{-1}y)\bar{\rho}_\lambda(dy) \\
&= \sum_{y_i}\lambda\ell(\lambda^{-2}t, y_i)\lambda^{1/\alpha}(V(y_i) - V(y_i-)) = \lambda^{(\alpha+1)/\alpha}\phi(\lambda^{-2}t)
\end{aligned} \tag{152}$$

and therefore its generalized inverse satisfies $\psi_\lambda(t) = \lambda^2\psi(\lambda^{-(\alpha+1)/\alpha}t)$. The rescaled singular diffusion defined by $\bar{Z}_\lambda = W_\lambda(\psi_\lambda(t))$ that has the same distribution as $\bar{Z}$ thus satisfies

$$\bar{Z}_\lambda(t) = W_\lambda(\psi_\lambda(t)) = \lambda\bar{Z}(\lambda^{-(\alpha+1)/\alpha}t). \tag{153}$$

Clearly, the triplet $(W_\lambda, V_\lambda, \bar{Z}_\lambda)$ has the same distribution as $(W, V, \bar{Z})$ too. We thus have

$$\bar{\mathbb{P}}(\bar{Z}(1) \in J_m^\eta(V)) = \bar{\mathbb{P}}(\bar{Z}_\lambda(1) \in J_m^\eta(V_\lambda)). \tag{154}$$

The set $J_m^\eta(V_\lambda)$ satisfies $J_m^\eta(V_\lambda) = \lambda J_{m\lambda^{-1}}^{\eta\lambda^{-1/\alpha}}(V)$ as can be easily verified from the scaling of $V$ or from (151) and thus

$$\begin{aligned}
\bar{\mathbb{P}}(\bar{Z}(1) \in J_m^\eta(V)) &= \bar{\mathbb{P}}(\lambda\bar{Z}(\lambda^{-(\alpha+1)/\alpha}) \in \lambda J_{m\lambda^{-1}}^{\eta\lambda^{-1/\alpha}}(V)) \\
&= \bar{\mathbb{P}}(\bar{Z}(\lambda^{-(\alpha+1)/\alpha}) \in J_{m\lambda^{-1}}^{\eta\lambda^{-1/\alpha}}(V)).
\end{aligned} \tag{155}$$

The proof is complete taking $\lambda$ satisfying $\lambda^{-(\alpha+1)/\alpha} = t$. $\quad\square$

PROOF OF LEMMA 9.7. The claim of the lemma is equivalent with

$$\int_0^1 \bar{\mathbb{P}}(\bar{Z}(t) \notin J_{m'}^{\eta'}) \, dt \le \delta'. \tag{156}$$

We use $\sigma(m)$ to denote the first time $\bar{Z}$ leaves $[-m, m]$. Let $m'$ be large enough such that

$$\bar{\mathbb{P}}(\sigma(m') < 1) < \delta'/2 \tag{157}$$

and let $\sigma = \sigma(m')$. Then

$$\begin{aligned}
\int_0^1 \bar{\mathbb{P}}(\bar{Z}(t) \notin J_{m'}^{\eta'}) \, dt &= \bar{\mathbb{E}}\left[\int_0^1 \mathbf{1}\{\bar{Z}(t) \notin J_{m'}^{\eta'}\} \, dt\right] \\
&\le \bar{\mathbb{E}}\left[\int_0^1 \mathbf{1}\{\bar{Z}(t) \notin J_{m'}^{\eta'}, \sigma \ge 1\} \, dt + \int_0^1 \mathbf{1}\{\sigma < 1\} \, dt\right] \\
&\le \bar{\mathbb{E}}\int_0^\sigma \mathbf{1}\{\bar{Z}(t) \notin J_{m'}^{\eta'}\} \, dt + \delta'/2.
\end{aligned} \tag{158}$$

We should bound the expectation in the last expression by $\delta'/2$. We establish this bound by proving

$$\bar{\mathbb{P}}\left[\bar{\mathbb{E}}\left(\int_0^\sigma \mathbf{1}\{\bar{Z}(t) \notin J_{m'}^{\eta'}\} \, dt \,\Big|\, V\right) \ge \delta'/4\right] \le \delta'/4. \tag{159}$$

The conditional expectation inside the brackets can be written as

$$\bar{\mathbb{E}}\left[\int_0^\sigma \mathbf{1}\{\bar{Z}(t) \notin J_{m'}^{\eta'}\} \, dt \,\Big|\, V\right] = \sum_{\substack{x_i \in [-m', m'] \\ v_i < \eta'}} G_{m'}(0, x_i) v_i, \tag{160}$$

where as usual $(x_i, v_i)$ is the collection of atoms of $\bar{\rho}$ and $G_m(x, y)$ is the Green function of the standard Brownian motion killed on exit from $[-m, m]$. There exists a constant $k$ depending only on $m$ such that $G_m(0, x) \le k$ for all $x \in [-m, m]$. We thus have

$$\bar{\mathbb{P}}\left[\bar{\mathbb{E}}\left(\int_0^\sigma \mathbf{1}\{\bar{Z}(t) \notin J_{m'}^{\eta'}\} \, dt \,\Big|\, V\right) \ge \delta'/4\right] \le \bar{\mathbb{P}}\left[k \sum_{\substack{x_i \in [-m', m'] \\ v_i < \eta'}} v_i \ge \delta'/4\right]. \tag{161}$$

The sum in the last equation has the same distribution as the Lévy process $V$ without jumps larger than $\eta'$ at the time $2m'$. One can thus easily choose $\eta'$ small enough, such that the last probability is smaller than $\delta'/4$. $\quad\square$

## REFERENCES

[1] Ben Arous, G. (2002). Aging and spin-glass dynamics. In *Proceedings of the ICM* **3** 3–14. MR1957514

[2] Ben Arous, G., Bovier, A. and Gayrard, V. (2002). Aging in the random energy model. *Phys. Rev. Lett.* **88** 087201.

[3] Ben Arous, G., Bovier, A. and Gayrard, V. (2003). Glauber dynamics of the random energy model, II. Aging below the critical temperature. *Comm. Math. Phys.* **235** 379–425. MR1974509

[4] Ben Arous, G., Černý, J. and Mountford, T. (2005). Aging for Bouchaud's model in dimension two. *Probab. Theory Related Fields.* To appear.

[5] Ben Arous, G., Dembo, A. and Guionnet, A. (2001). Aging of spherical spin glasses. *Probab. Theory Related Fields* **120** 1–67. MR1856194

[6] Bouchaud, J. P., Cugliandolo, L., Kurchan, J. and Mézard, M. (1998). Out-of-equilibrium dynamics in spin-glasses and other glassy systems. In *Spin-Glasses and Random Fields* (A. P. Young, ed.). World Scientific, Singapore.

[7] Dembo, A., Guionnet, A. and Zeitouni, O. (2001). Aging properties of Sinai's random walk in random environment. Preprint. Available at http://arxiv.org/abs/math/0105215.

[8] Fontes, L. R. G., Isopi, M. and Newman, C. M. (1999). Chaotic time dependence in a disordered spin system. *Probab. Theory Related Fields* **115** 417–443. MR1725402

[9] Fontes, L. R. G., Isopi, M. and Newman, C. M. (2002). Random walks with strongly inhomogeneous rates and singular diffusions: Convergence, localization and aging in one dimension. *Ann. Probab.* **30** 579–604. MR1905852

[10] Mathieu, P. (2000). Convergence to equilibrium for spin glasses. *Comm. Math. Phys.* **215** 57–68. MR1799876

[11] Rinn, B., Maass, P. and Bouchaud, J.-P. (2000). Multiple scaling regimes in simple aging models. *Phys. Rev. Lett.* **84** 5403–5406.

[12] Stone, C. (1963). Limit theorems for random walks, birth and death processes, and diffusion processes. *Illinois J. Math.* **7** 638–660. MR158440

Courant Institute
  of Mathematical Sciences
New York University
251 Mercer Street
New York, New York 10012-1185
USA
e-mail: benarous@cims.nyu.edu

Weierstrass Institute for
  Applied Analysis and Stochastics
Mohrenstr. 39
10117 Berlin
Germany
e-mail: cerny@wias-berlin.de